\documentclass[graybox]{svmult}

\usepackage{type1cm}        % activate if the above 3 fonts are
\usepackage{makeidx}         % allows index generation
\usepackage{graphicx}        % standard LaTeX graphics tool
\usepackage{cite}         
\usepackage{multicol}        % used for the two-column index
\usepackage[bottom]{footmisc}% places footnotes at page bottom

\usepackage{newtxtext}       % 
\usepackage[varvw]{newtxmath}       % selects Times Roman as basic font

\usepackage{apj-jour}

\makeindex             % used for the subject index
\def \astena{{ASTENA}\/}

\def \integral{{\em INTEGRAL}\/}

\def \eastrogam{{\em e-Astrogam}\/}

\def \phcmsec{\hbox{photons$\,$cm$^{-2}$s$^{-1}$} }

\begin{document}

\title*{Hard X-ray/Soft gamma-ray Laue Lenses for High Energy Astrophysics}
\author{Filippo Frontera}
\institute{Filippo Frontera \at University of Ferrara, Department of Physics and Earth Sciences, Via Saragat, 1 - 44122 Ferrara, and INAF OAS, Via Gobetti 101, 40129 Bologna; \email{filippo.frontera@unife.it}}
%\and Name of Second Author \at Name, Address of Institute \email{name@email.address}}
\maketitle

\abstract{The study of the celestial phenomena in the hard X-ray/soft gamma-ray band(20 keV--1 MeV) is very intriguing but also very difficult to be performed with the needed sensitivity. In this review I will discuss the astrophysical importance of the soft gamma-ray astronomy, its difficulties to solve its issues with the current instrumentation, and a possible solution achievable  using focusing Laue lens. Concerning these instruments, I will discuss their functioning principle, how to achieve a high  reflection efficiency, their imaging properties, the current feasibility studies, the technological developments and observation prospects.}

\section{Introduction}
\label{s:intro}
Hard X-ray/soft gamma-ray astronomy is a crucial  window for the study of very energetic and violent events in the Universe.  With the ESA INTEGRAL observatory \cite{Winkler03}, and the NASA {\em Swift} satellite \cite{Gehrels04}, unprecedented sky surveys in the band beyond 20 keV are being performed \cite{Krivonos22,Oh18}.  As a consequence, thousands of celestial hard X-ray sources have already been discovered, new classes of Galactic sources are identified, an unprecedented overview of the extragalactic sky is available \cite{Malizia23}, with several issues still open about, e.g., the soft gamma-ray spectra of Soft Gamma Ray Repeaters and anomalous X--ray pulsars (see, e.g., \cite{Gotz06,Kuiper06}), a deep study of high energy emission physics in AGNs, the origin of Cosmic hard X/soft gamma-ray diffuse background (see, e.g., \cite{Ajello09,Horiuchi10}).

\begin{figure}
\centering
\includegraphics[width=0.70\textwidth]{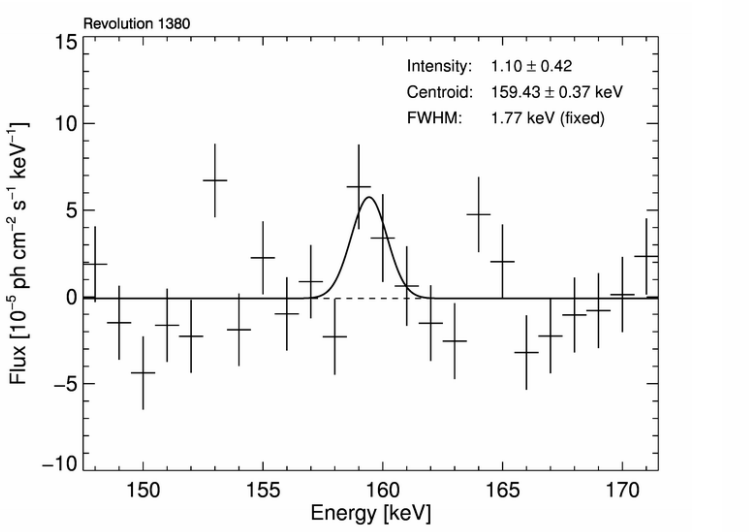}
\caption{The 158 keV line due to the radioactive $^{56}$Ni decay of the type Ia SN2014J, as observed with 150.24 ks \integral\ SPI observations, 3 weeks after the explosion. The line intensity is given in $10^{-4}$~\phcmsec\ units. Reprinted from \cite{Diehl14}.}
     \label{f:158keV-line}
 \end{figure}

In addition an extended matter-antimatter annihilation emission from the Galactic  center region is observed (see, e.g., \cite{Weidenspointner08,Siegert16a}, and Galactic nucleosynthesis processes have also been reported \cite{Weidenspointner08,Diehl06}. However, concerning the annihilation line, its origin is still a mystery. A possible origin from low mass X-ray binaries with strong emission at hard X--ray photon energies (hard LMXBs) was proposed by \cite{Weidenspointner08}, but, due to the low angular resolution of the best soft gamma-ray instruments of the current generation (about 3 degrees  in the case of SPI), this hypothesis cannot be tested. 
Concerning the nucleosynthesis processes, with the \integral\ SPI instrument it was possible to find a weak evidence of the 158 keV line from the  radioactive $^{56}Ni$  produced in a type I supernova explosion \cite{Diehl14} (see Fig.~\ref{f:158keV-line}), but it is clear that a more sensitive study, and, for mapping the line across the SN surface, a much better angular resolution, are needed.

Furthermore, with the discovery of the X-ray afterglow of Gamma Ray Bursts (GRBs) \cite{Costa97} and the consequent discovery of their extragalactic origin \cite{Metzger97}, the knowledge of the broad band spectral properties of their afterglow, in particular the soft gamma--ray afterglow spectrum, is needed to better understand the afterglow origin and its emission mechanism (see, e.g., \cite{Corsi05, Kouveliotou13}). 

From the examples above, it is apparent that a new generation of soft gamma--ray telescopes is needed.
The current generation  has relied on the use sky direct--viewing detectors with mechanical collimators (see, e.g., BeppoSAX/PDS, \cite{Frontera97}) and, in some cases, with modulating aperture systems, like coded masks (e.g., INTEGRAL/IBIS, \cite{Ubertini03}).
These telescopes have  modest sensitivities that improve at best as the square root of the detector surface.  

The only solution to the limitations of the current generation of gamma--ray instruments is the use of focusing instruments. As we will see, Laue lenses, based on diffraction from crystals in transmission configuration, can do that in the hard X--/soft gamma--ray ($<$ 1 MeV) domain and thus can provide a big leap in flux sensitivity with respect to the best non-focusing instruments. In addition, they can significantly improve the 
angular resolution with respect to mask telescopes (e.g., INTEGRAL IBIS). 

% Figure 1
%
%---------------------------------------------------------------- 
%-----------------------------------------
%

In this paper, I will review the basic properties of Laue lenses, their optimization criteria, the current development status and the prospects for future missions devoted to  soft gamma--ray astronomy.

\section{Laue lens concept}

Laue lenses exploit the interference between the periodic nature of light and a~periodic structure such as the matter in a~crystal.  
In order to better understand this interference, basic elements of crystallography are needed, for which I refer to text books like \cite{Zachariasen45,Cullity78}. 

I limit here to refer to the periodic structure of a crystal  with a basic pattern (or unit cell), with one or more atoms or molecules (point lattices), which is periodically repeated.
%(see Fig.~\ref{f:unit-cell}).  
%
% Figure 2
%
%\begin{figure}[htbp]
%	\begin{center}
%		\includegraphics[width=0.6\textwidth]{Figures/fig-unit-cell.png}
%   		\caption{The unit cell of a crystal, with axes given by the vectors {\bf a, b, c}, and angles $\alpha$, $\beta$, $\gamma$ between these axes. When the vectors are  all equal and perpendicular with each other, the cell is cubic.}
%		\label{f:unit-cell}
%	\end{center}
%\end{figure}

The orientation of planes in a lattice  can be represented through the Miller indices $(hkl)$, whose reciprocals give the fractional intercepts which the chosen lattice plane makes with the crystallographic axes $a$, $b$, and $c$. 
%see Fig.~\ref{f:unit-cell}). 
The spacing between adjacent lattice planes $(hkl)$ is given by $d_{hkl}$, while the volume  of the unit cell is given by $V$.
For a cubic cell with side length $a$:
%
% Equation 2
%
\begin{eqnarray}
\begin{aligned}
d_{hkl} &= \frac{a}{h^2+k^2+l^2} \qquad
V = a^3
\end{aligned}
\label{e:dhkl-and-V}
\end{eqnarray}

%For other cell geometries see literature, e.g., \cite{Cullity78}.
%
%
\subsection{The Bragg law}

In order to be diffracted, an incoming gamma-ray photon must satisfy the Bragg--law (see Fig.~\ref{f:Bragg}), that relates the spacing between crystal planes $d_{hkl}$ and the incidence angle $\theta_B$ to the photon wavelength $\lambda$ and, alternatively, to its energy $E$:
%
% Figure 2
%
\begin{figure}[htbp]
	\begin{center}
		\includegraphics[width=0.45\textwidth]{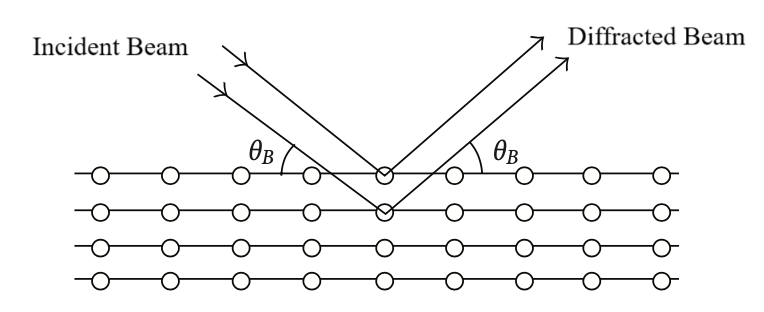}
        \includegraphics[width=0.45\textwidth]{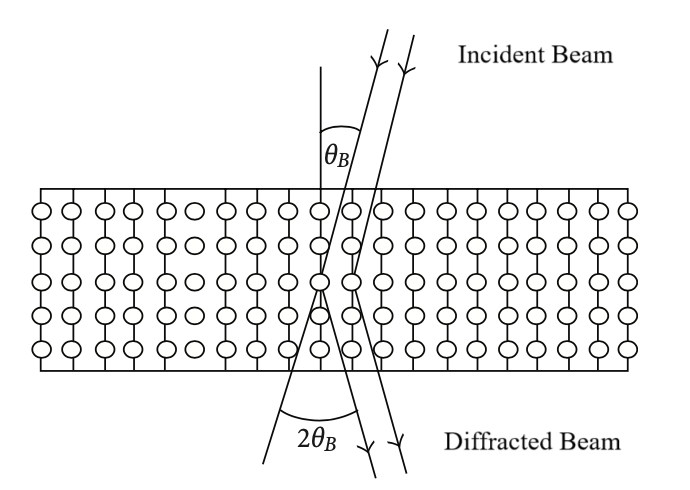}		%\includegraphics{Bragg_principle.pdf}
   		\caption{The Bragg condition for constructive interference of a gamma-ray photon with the atoms of a given crystalline plane. {\em Left:} Diffraction in reflection configuration also known as Bragg geometry. {\em Right:} Diffraction in transmission configuration, also known as Laue geometry.}
		\label{f:Bragg}
	\end{center}
\end{figure}
%
% Equation 3
%
\begin{equation}
2 d_{hkl} \sin \theta_B  = n \lambda= n \frac{hc}{E}
\label{e:bragg}
\end{equation}

where $n$ is the diffraction order and $hc = 12.4$~keV$\cdot$\AA, with $d_{hkl}$ and $\lambda$ in \AA\ and $E$ in keV units.

As shown in Fig.~\ref{f:Bragg}, two crystal configurations are possible: reflection configuration (also called Bragg geometry, see left panel) and transmission configuration (also called Laue geometry, see right panel).

\subsection{Geometry of a Laue lens}

\begin{figure}[ht]
	\begin{center}  	
	\includegraphics[width=0.5\textwidth]{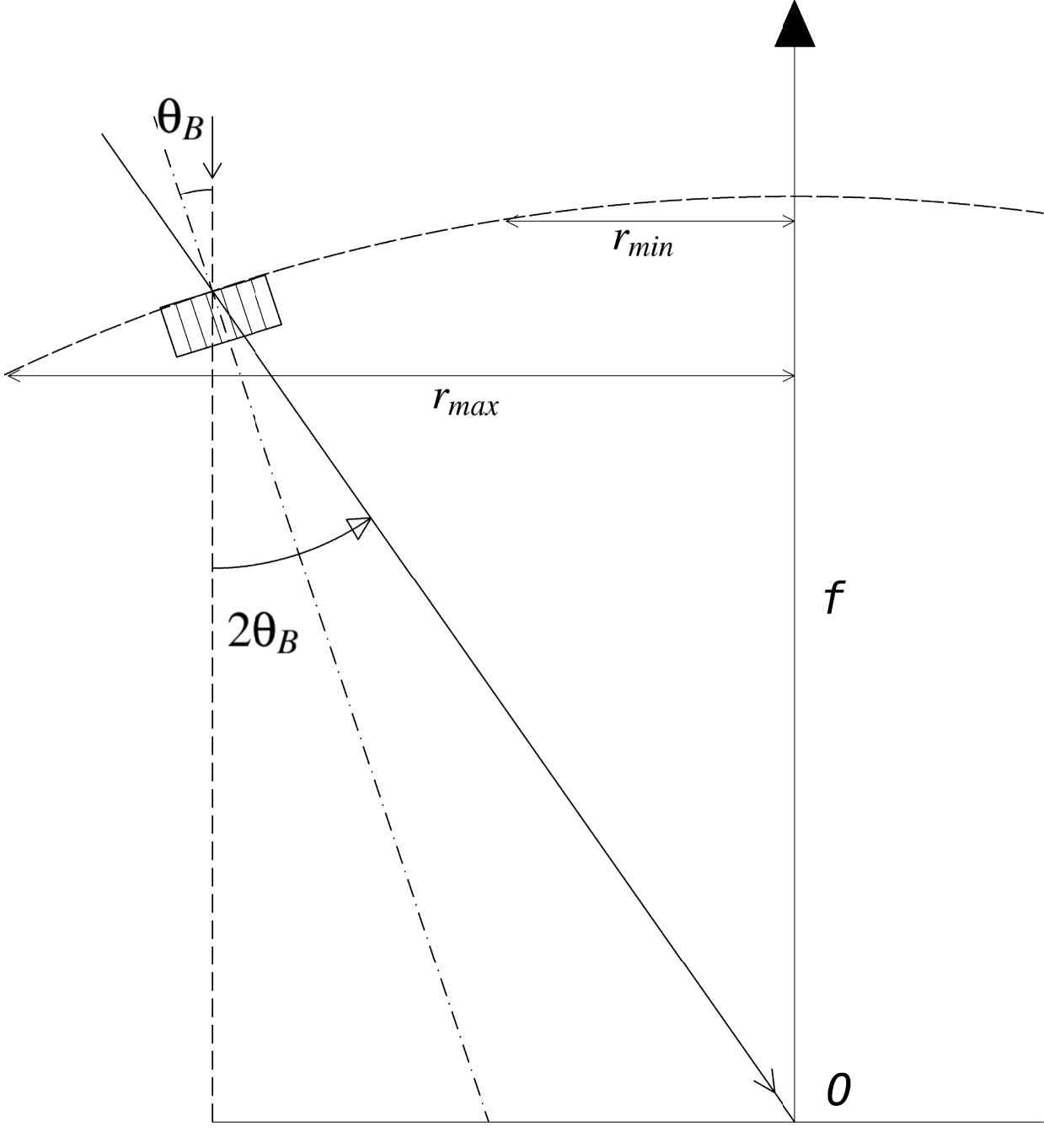}
		\caption{Laue lens geometry. The lens focal length is given by $f = R/2$ where $R$ is the radius of the spherical cup, and the diffracting crystals on the cup are oriented in such a way to focus in the point {\bf O} (lens focus) the soft gamma--ray photons that, in the case of a lens for astrophysical applications, are coming from infinity.}
		\label{f:lens-geometry}
	\end{center}
\end{figure}

A Laue lens  is made of a large number of crystal tiles with diffracting planes in transmission configuration (see Fig.~\ref{f:Bragg}), disposed on a spherical cup of radius $R$, with their diffracting planes perpendicular to the sphere (spherical geometry, see Fig.~\ref{f:lens-geometry}). The focal spot is in the point $0$ of the symmetry axis, at a distance $f = R/2$ from the cup, with  $f$ being called {\em focal length} (see below).  Eq.~\ref{e:bragg} requires that each crystal has to be oriented in such a way that the angle between the incident beam and the crystalline planes is given by the Bragg angle~$\theta_B$. Thus, in order to focus a photon with energy $E$, for the first order diffraction which is the most efficient, the crystal has to be at a distance $r$ from the symmetry axis given by  

%
% Equation 4
%
\begin{equation}
r = f \sin 2\theta_B \approx f \frac{hc}{d_{hkl}} \frac{1}{E}
\label{e:lens-radii}
\end{equation}

with $d_{hkl}$ in \AA\ and $E$ in keV units, where the approximated expression is valid for gamma--ray lenses, given the small diffraction angles involved and thus $cos \theta_B \approx 1$,  

%
% Figure 3
%
From equation~\ref{e:lens-radii}, it can be derived the energy $E$ of the incident photons that, at given distance from the lens symmetry axis, can be reflected by the lens in its focus:
%
% Equation 5
%
\begin{equation}
E \approx f\, \frac{hc}{d_{hkl}}\,\frac{1}{r}
\label{e:lens-energy}
\end{equation}
\subsection{Lens energy bandpass, inner and outer radii and role of the focal length}

Any Laue lens can be designed in such a way to diffract photons over a prefixed energy passband ($E_{min}$,$E_{max}$). From eq.~\ref{e:lens-energy}, at the first order diffraction, which is the most efficient, it is possible to get the energy passband and, thus the inner and outer radii of a lens ($r_{min}$, $r_{max}$) needed to get the requested lens passband:
%%
% Equation 6
%
%\begin{eqnarray}
%\begin{aligned}
%E_{min} \approx \frac{hc f}{d_{hkl}~r_{max}} \\ \\
%E_{max} \approx \frac{hc f}{d_{hkl}~r_{min}}.
%\end{aligned}
%\label{eq:Emin-Emax}
%\end{eqnarray}

%
%
% Equation 6
%
\begin{eqnarray}
\begin{aligned}
r_{min} \approx \frac{hc f}{d_{hkl}~E_{max}}  \qquad
r_{max} \approx \frac{hc f}{d_{hkl}~E_{min}}.
\end{aligned}
\label{e:rmin-rmax}
\end{eqnarray}
The focal length $f$ is crucial to extend the lens passband. Given that high energy photons are focused by the innermost part of the lens, the inner lens area can be increased only increasing the focal length 
(the lens area approximately increases with $f^2$). 

\subsection{Crystal tile distribution in the lenses and their focusing}
\label{crystal-focusing}

\begin{figure}
\begin{center}
\includegraphics[angle=0, width=0.45\textwidth]{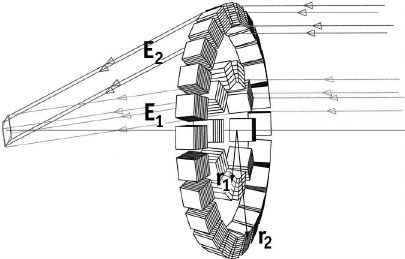}
\includegraphics[angle=0, width=0.45\textwidth]{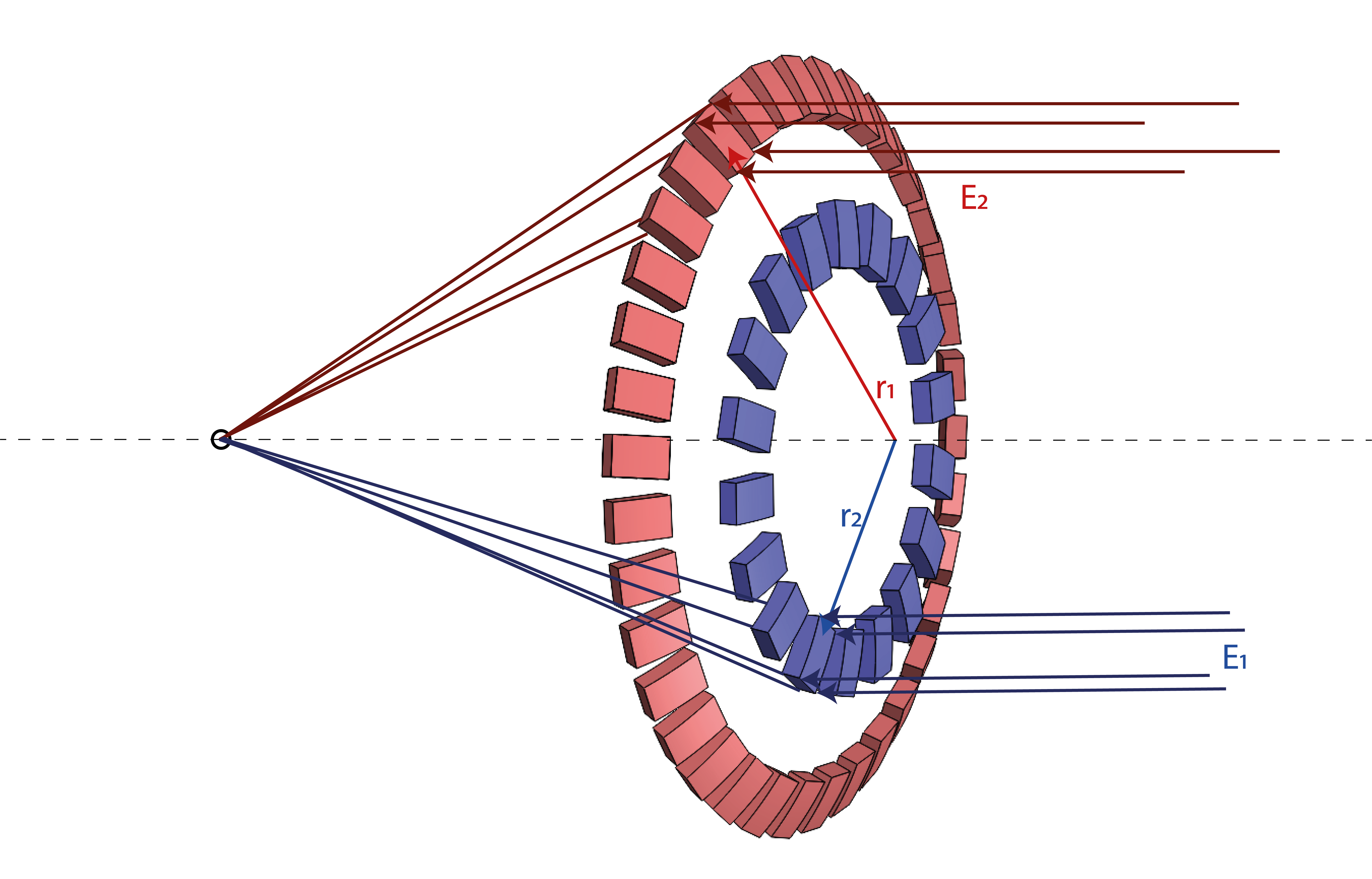}
\end{center}
%\vspace{-0.5cm}
\caption{Distribution, in concentric rings, of the crystals in a Laue lens with high focal length ($>10$~m). If the crystal tiles, for a ring of radius $r$, have the same diffracting planes ($hkl$), all of them reflect in the focus photons of energy $E$ given by Eq.~\ref{e:lens-energy}. {\em Left}: focusing of  mosaic crystals. Figure reprinted from \cite{vonBallmoos05}. {\em Right}: focusing of bent crystals.
	%{\em Right}:  crystal tiles disposed along an Archimedes' spiral results in a 
	%continuously varying energy $E$. Given the footprint of the crystals, the image in the 
	%focal plane has as minimum size that of the crystal size.
 }
\label{f:crystal-tileDistribution}
\end{figure}

 Given the spherical geometry of the lens, assuming that the crystal material and the diffracting planes ($hkl$) are the same for the entire lens, the best distribution of the crystal tiles on the spherical cup, for a high focal length ($>10$~m), is on concentric rings (see Fig.~\ref{f:crystal-tileDistribution}). Thus, for a given $r$, the diffracted energy $E$ is given by Eq.~\ref{e:lens-energy} for all crystals in each ring. 
For Laue lenses with small focal length ($<$ 10 m), in order to optimize the smoothness of the lens effective area versus energy (see below), the best geometrical distribution of the crystals on the spherical cup would be an Archimedes spiral (see, e.g., \cite{Pellicciotta06}). 

In the case of flat diffracting planes perpendicular to the crystal cross section, like perfect crystals in Laue geometry, the reflected photon distribution in the focal plane have as the same size of the cross section of the crystal tiles. A similar distribution of the reflected photons in the focal plane is obtained in the case of a special class of flat diffracting planes called mosaic crystals (see next subsection), which are made of many microscopic perfect crystals (\textit{crystallites}) with their flat diffracting  planes $(hkl)$ slightly misaligned with each other around a mean direction, called \textit{mean lattice plane}). In the left panel of Fig.~\ref{f:crystal-tileDistribution} it is shown the focusing in the case of mosaic crystals. 

The best focusing of a lens is obtained in the case of bent crystals with a curvature radius equal to that of the sphere (see right panel of Fig.~\ref{f:crystal-tileDistribution}).
%while a uniformly changing value of $r$ gives rise to an Archimedes spiral (right panel of Fig.~\ref{f:LaueLensPrinciple}).  
%in the first case (constant $r$) the energy of the diffracted photon will be centered on~$E$ for 
%all the crystals in the ring, while in the second case (Archimedes spiral), 
%the reflected energy $E$ will continuously vary from one crystal to the other.
%

%
% Figure 4
%

\subsection{Effective area of a Laue lens}

The effective area at energy $E$ of a focusing telescope, like the X-ray optics based on total reflection, is defined as the geometrical collecting area $A_C$ of the telescope projected in the focal plane  times the reflection efficiency at energy $E$. This definition is also valid for Laue lenses, but, in this case, the collecting area depends on the energy bandwidth $\Delta E$, centred on $E$, i.e., $A_C = A_C (\Delta E)$. 

Given that, specially for astronomical applications, the lens effective area in the entire lens passband is desired to be covered in a smooth manner as a function of energy, the energy response of single or contiguous crystals has to superimpose with each other. This condition has an impact on the  angular bandwidth of the crystals suitable for Laue lenses.
Indeed, from the derivative of the Bragg relation in the small
angle approximation which is our case ($2\mathrm{d}\theta_{\mathrm{B}} \approx
\mathrm{h}\mathrm{c}/\mathrm{E}$), we can get:
\begin{equation}
  \Delta E/E = \Delta \theta_B/\theta_B
    \label{eq:angular width}  
\end{equation}
where $\mathrm{\Delta}\theta$ is the angular width of the crystal. Thus the energy bandwidth $\mathrm{\Delta}\mathrm{E}$ of a crystal, for diffraction, is given by

\begin{equation}
\Delta E= \frac{2d \, E^2 \, \Delta \theta_B}{n \, h \, c}.
\label{e:DeltaE}
\end{equation}

%It is worth pointing out that, whereas the energy bandpass of a~crystal grows with the square of
%energy, Doppler broadening of astrophysical lines (e.g. in SN ejecta)
%increases linearly with energy for a~given expansion velocity.

Since the acceptance angle  $\Delta \theta$  of perfect crystals (known as {\em Darwin width}) is extremely narrow (fractions of arcsec to a few arcsec (see, e.g., \cite{Zac45}), such crystals are not suitable for  Laue lenses. Instead, it is possible to use either mosaic crystals or bent crystals.

Mosaic crystals, thanks to the angular distribution of their crystallites, also called \textit{mosaicity}, can cover in a smooth manner the effective area of the lens with energy. In addition (see next section) their flux throughput increases with mosaicity, but at expenses of the angular resolution of the lens. 
%A good compromise is to use mosaic crystals with angular widths from a~few tens of arc seconds to a~few arc minutes, that are acceptable for lenses in which a high angular resolution is not required. 
Instead bent crystals with curvature radius equal to that of the lens, naturally have the angular width that provide the best focusing of the diffracted photons.

However, in both cases, for a lens it is important not only the focusing features, but also the reflectivity of the crystals adopted.

\section{Mosaic crystals and their reflectivity}
\label{s:mosaic-refl}

As mentioned above, mosaic crystals are made of \textit{crystallites}, with the lattice planes $(hkl)$, chosen for diffraction, slightly misaligned with each other around the \textit{mean lattice plane}.
The distribution function of the crystallite misalignments from the mean lattice plane can be approximated with a Gaussian function:
\begin{equation}
\label{eq:Gauss}
W(\delta)=\frac{1}{\sqrt{2\pi}\eta}
\exp{\left( - \frac{\delta^2}{2\eta^2}\right )} \, ,
\end{equation}
where $\delta$ is the magnitude of the angular deviation of the crystallites from the mean lattice plane, while 
$\beta = 2.35 \eta$ is the fwhm of the mosaic spread ({\em mosaicity}). 

The reflectivity $R(\delta, E)$ of mosaic crystals configured in a lens  as in Fig.~\ref{f:lens-geometry}, with a parallel incident beam, is  given by \cite{Zac45}:

\begin{equation}
R(\delta, E) = \frac {I_d (\delta, E)}{I_0} = \sinh{(\sigma T)}
\exp{\left [ - \left (\mu + \gamma _0 \sigma \right )
\frac {T}{\gamma _0} \right ]} = 
\frac{1}{2}(1- e^{-2\sigma T}) e^{ - \mu \frac {T}{\gamma _0}}
\, 
\label{e:refl_mosaic}
\end{equation}
\\
\\
where $I_0$ is the intensity of the incident beam, $\mu$ is the total absorption coefficient per unit of length corresponding to that energy, $\gamma_0$ is the cosine of the angle between the direction of the photons and the normal to 
the surface, $T$ is the thickness of the mosaic crystal and $\sigma$ is given by: 
\begin{equation}
\label{eq:sigma}
\sigma = \sigma (E, \delta) = W(\delta) Q(E) f(A) \, ,
\end{equation}
where
\begin{equation}
Q (E) = \left | \frac {r_e^2 F_{hkl}}{V}\right |^2 \,
\lambda ^3 \, \frac {1+\cos^2(2 \theta_B)}{2 \sin 2\theta_B} \, ,
\label{eq:q}
\end{equation}
in which $r_e$ is the classical electron radius, $F_{hkl}$ is the structure factor of the crystal diffracting planes, inclusive of the temperature effect (Debye-Waller's factor), 
$V$ is the volume of the crystal unit cell, $\lambda$ is the radiation wavelength 
and $\theta_B$ is the Bragg angle for that particular wavelength, 
while $f(A)$ is given by:
%
% Equation
%
\begin{equation}
f(A)= \frac{B_0(2A) + |\cos2\theta_B| \,  B_0(2A|\cos2\theta_B|)}{2A(1+ \cos^2\theta_B)} \\
\end{equation}
where $B_0$ is the Bessel function of zero order, integrated between 0 and $2A$, with 
$A$ defined as follows:
 \begin{equation}
 A = \frac{\pi \, t_0}{\Lambda_0 \, \cos \theta_B},
 \end{equation}
in which $t_0$ is the crystallite thickness, and $\Lambda_0$ ({\em extinction 
length}) is defined, for the symmetrical Laue case which is our case (see e.g., \cite{Authier01}), as:
\begin{equation}
\Lambda_0 = \frac{\pi\, V_c \, \cos \theta_B}{r_e \lambda \, |F_{hkl}| \, (1+ |\cos 2\theta_B|},
\label{e:extin_length}
\end{equation} 
for unpolarised radiation. More in general (see \cite{Authier01}):
\begin{equation}
\Lambda_0 = \frac{\pi\, V_c \, \cos \theta_B}{r_e \lambda \, |F_{hkl}| \, C},
\label{e:extin_length1}
\end{equation} 
where $C$ is the polarization factor of the incident radiation.

In general $f(A)<1$ and converges to 1 for $t_0 \ll \Lambda_0$. In this 
case we get the highest reflectivity.

The quantity $\gamma_0\sigma$ is the {\em secondary extinction} 
coefficient and $T/\gamma_0$ is the 
distance travelled by the direct beam inside the crystal.

As can be seen from  Eq.~\ref{e:refl_mosaic}, in spite of the best choice of crystal material (see Section~\ref{s:cryst-mat}), its diffracting planes, thickness and so on, the maximum reflectivity of a mosaic crystal is 50\%. If we want to get a higher reflectivity, a different solution for Laue lens crystals has to be found.

\subsection{Bent crystals and their reflectivity}
\label{s:bent_cryst}

As mentioned above, bent crystals with curvature radius  equal to that of the lens surface,  have the advantage of giving the best focusing of the reflected photons and a very smooth dependence of the effective area with energy.

In addition to these properties,  the primary curvature  produced from the crystal bending, can generate a curvature in the diffracting planes perpendicular to the crystal cross section. For particular crystallographic planes; in the other cases the diffracting planes, this effect is particularly important for the diffraction efficiency (ratio between the diffracted and the transmitted photons) and thus for the crystal reflectivity. 

\begin{figure}
\begin{center}
\includegraphics[angle=0, width=0.90\textwidth]{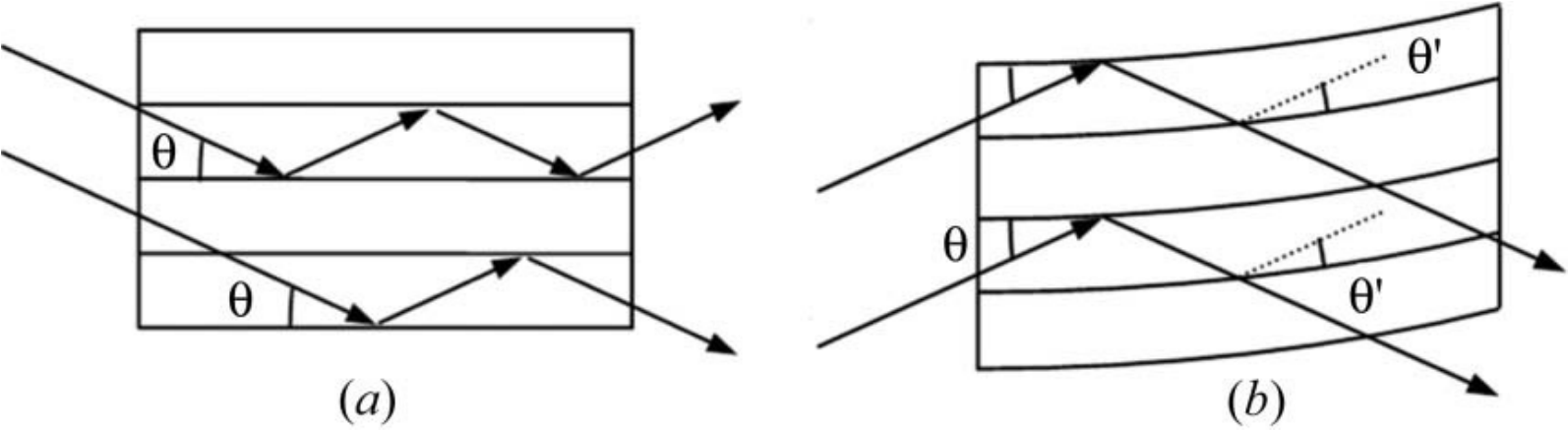}
\end{center}
%\vspace{-0.5cm}
\caption{X--ray diffraction in Laue geometry for flat (a) and bent (b) diffracting planes. Reprinted from \cite{Bellucci13}}
\label{f:fdp-vs-cdp}
\end{figure}

Indeed, the formation of Curved Diffracting Planes (CDP) makes a single diffraction in these planes very likely and thus the reflectivity is not limited to a maximum of 50\% as in the case of flat diffracting planes (see Fig.~\ref{f:fdp-vs-cdp}), but it can achieve values close to unity \cite{Keitel99,Malgrange02} (see also below). The formation of CDP crystals is also apparent in an increased energy bandwidth of these crystals (a few arcsec), for which they are also called quasi-mosaic (QM) crystals.    

%
% Figure 5
%

%CDP crystals can also be obtained by applying a proper thermal gradient to the crystal sides (see, e.g., \cite{Smither05}), 2) by growing a two-component crystal, like Si$_{1-x}$Ge$_x$ (see, e.g., \cite{Keitel99,Abrosimov05b}, 3) by bending the crystal, technique firstly adopted in synchrotron X--ray beamlines, obtained by applying an external bending moment to the crystal. 
%More recently bending technologies have been developed to obtain a self-standing curvatures, important for their use in Laue lenses. Bent Self-standing crystals have been obtained, e.g., by lapping thin ($<$2 mm) crystals \cite{Ferrari13}, by performing indentations in the surface of thin ($<$2 mm) crystals (e.g., \cite{Barriere10,Bellucci11}), by depositing carbon fiber onto mono-crystals \cite{Camattari14}.

 One of the first investigations of the QM crystals was performed by \cite{Ivanov05} and was demonstrated that quasi-mosaicity  is an effect driven by crystalline anisotropy. 

Given the importance of these crystals for a Laue lens for increasing the crystal reflection efficiency and  the smoothness of the effective area with energy, we limit the discussion of the CDP reflectivity to these crystals. 

%
% Figure 6
%
\begin{figure}[ht]
\begin{center}
\includegraphics[angle=0, width=0.90\textwidth]{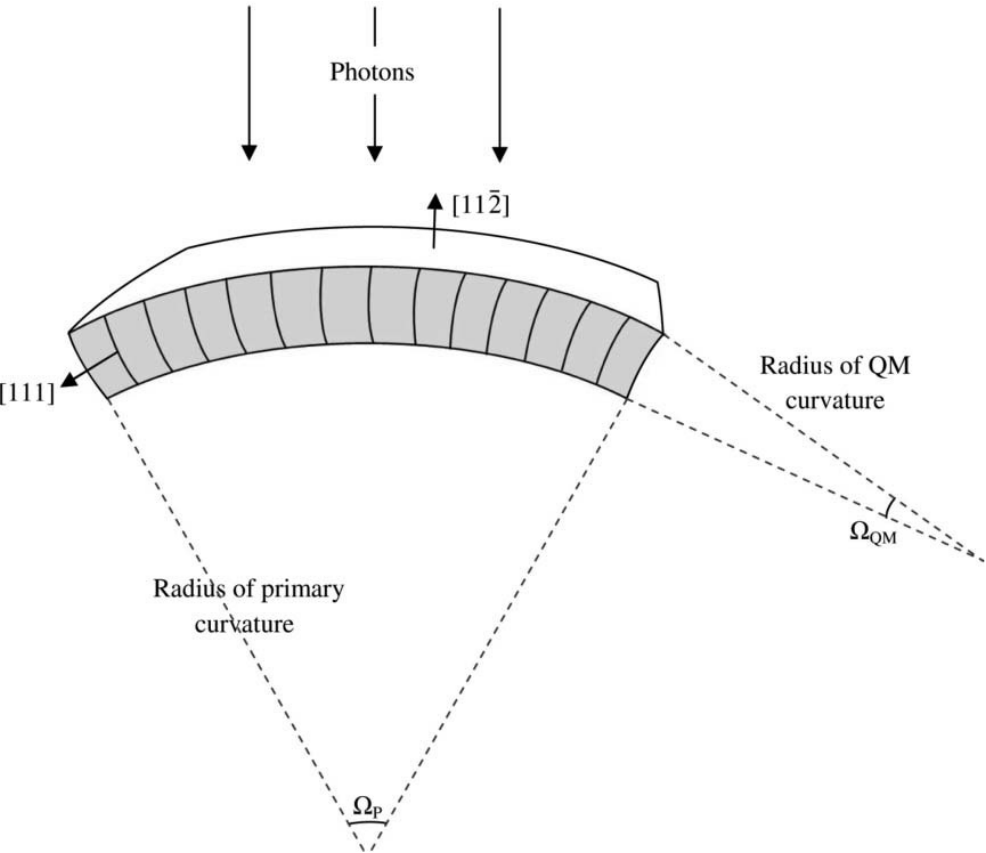}
\end{center}
%\vspace{-0.5cm}
\caption{Primary $\Omega_p$ and secondary or quasi-mosaic $\Omega_s$ bending angles in the case of (111) reflecting planes. The corresponding curvature radii are given by $R_p$ (lens curvature radius) and $R_s$ (secondary curvature radius). Reprinted from \cite{Guidi11}.}
\label{f:sec-vs-primary}
\end{figure}

The theory of the radiation diffraction in QM crystals, 
for the case of a homogeneous curvature, is due to \cite{Malgrange02,Authier01}. 
In this theory, the distortion of diffracting planes is described 
by a strain gradient $\beta_s$, that, in the case of a uniform curvature, is given
by:
\begin{equation}
\beta_s = \frac{2 \Omega_{QM}}{T_0 \delta_w}
\label{e:beta_s}
\end{equation}
where $\Omega_{QM}$ is the total bending angle of the diffracting planes, which is related to the quasi-mosaic curvature radius (see Fig.~\ref{f:sec-vs-primary}) $R_{QM}=T_0/\Omega_{QM}$, where
$T_0$ is the thickness of the crystal and $\delta_w$ is the Darwin width \cite{Zac45}.  

When the strain gradient $\beta_s$ becomes larger than a critical value $\beta_c = \pi /(2 \Lambda_0)$,
with extinction length $\Lambda_0$ given by Eq.~\ref{e:extin_length1}, the peak reflectivity $R^{peak}$ of the QM crystal is given by
\begin{equation}
R^{peak}(c_{QM}, E) = \frac{I_r^{peak} (c_{QM}, E)}{I_0}  = \left( 1 - e^{-{\pi^2 d_{hkl} \over c_{QM} \, \Lambda_0^2}} \right) \, e^{-{\mu \, \Omega_{QM} \over c_{QM} \, \cos \theta_B}}.
\label{e:refl_curved}
\end{equation}
where $c_{QM} = \Omega_{QM}/T_0$ is the curvature of the lattice planes, also called QM curvature, that is  assumed to be uniform
across the crystal thickness. 

\begin{figure}
\begin{center}
\includegraphics[angle=0, width=0.60\textwidth]{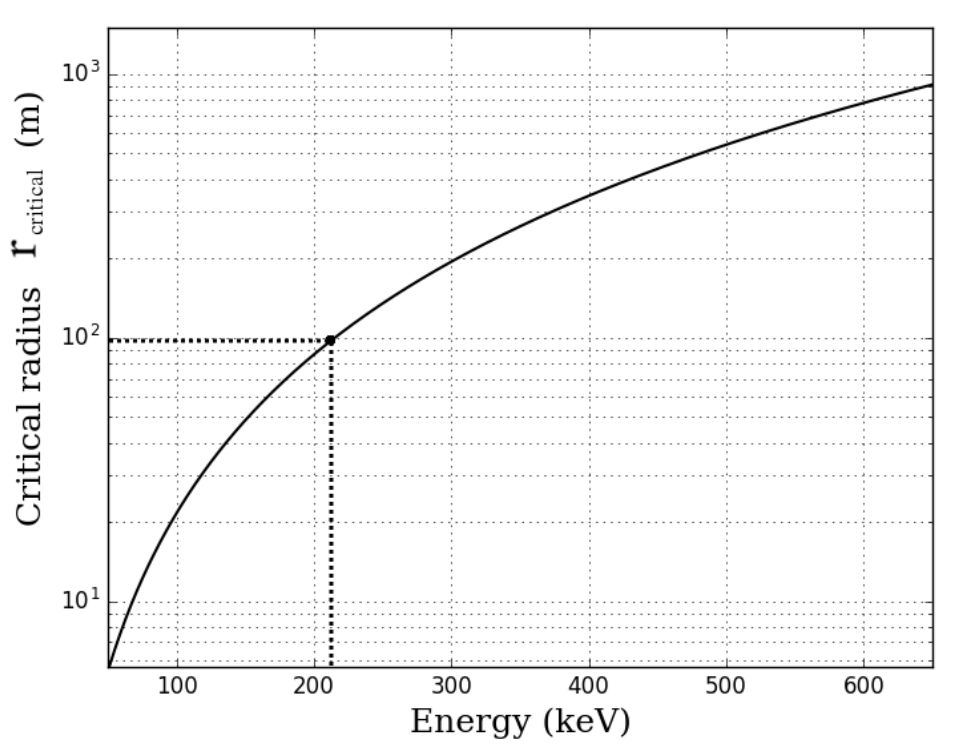}
\end{center}
%\vspace{-0.5cm}
\caption{Critical radius $R_c$ of the quasi-mosaic theory versus photon energy for Ge(111). It can be seen that the critical radius increases with energy. It is also shown (dotted lines) the secondary curvature radius of Ge(111) for a crystal primary curvature radius of 40 m. In this case, photon energies higher than about 200 keV satisfy the condition for the validity of the eq.~\ref{e:refl_curved}. Figure reprinted from \cite{Virgilli17}.}
\label{f:critical radius-vs-energy}
\end{figure}

It is important to note that, from the definition of $\beta_s$ (eq.~\ref{e:beta_s})  and $\beta_c$, the condition:
\begin{equation}
    \beta_s>\beta_c
\end{equation}
required for the validity of eq.~\ref{e:refl_curved}, implies that the secondary curvature radius $R_s$ has to be lower than a critical radius $R_c$, i.e., 
\begin{equation}
R_s < R_c = \frac{2\, \Lambda^2}{\pi \, d_{hkl}}
\label{e:R_c}
\end{equation}
with $R_c$ energy dependent through $\Lambda$. Thus, for a given bent crystal with quasi-mosaic structure of the diffracting planes, it is possible to establish the energy range  in which the reflectivity is given by eq.~\ref{e:refl_curved}. A plot of the critical radius, in the case of bent Ge(111), is shown in Fig.~\ref{f:critical radius-vs-energy} %
%
% Figure 7
%

For values of $R_s>R_c$, the CDP crystal reflectivity been evaluated by \cite{Bellucci13} using a multi-lamellar approach for the crystal structure, with the condition that the diffraction efficiency converges to 0.5 when $R_s \gg R_c$. 

The ratio between secondary and primary curvature radii is discussed by \cite{Guidi11}, in the framework of the linear elasticity theory.

\subsection{Techniques adopted for obtaining CDP crystals}
\label{s:techniques-for-CDP}

CDP crystals can be obtained in various ways.
One technique is the growing of a two-component crystal whose composition varies along the crystal growth axis, e.g., Si$_{1-x}$Ge$_x$ \cite{Keitel99,Abrosimov05b}. But the best techniques for Laue lenses, that provide also the best focusing, include the elastic bending of a perfect crystal tile (technique commonly adopted in synchrotron radiation facilities), the grooving of one of the surfaces of a crystal (working for thin ($\leq 2$~mm) crystals) \cite{Bellucci11}, the lapping of thin ($<$2 mm) crystals \cite{Ferrari13}, the deposition of a thin layer of carbon fiber onto mono-crystals \cite{Camattari14}.

%
% Figure 8
%
\begin{figure}[ht]
\begin{center}
\includegraphics[angle=0, width=0.49\textwidth]{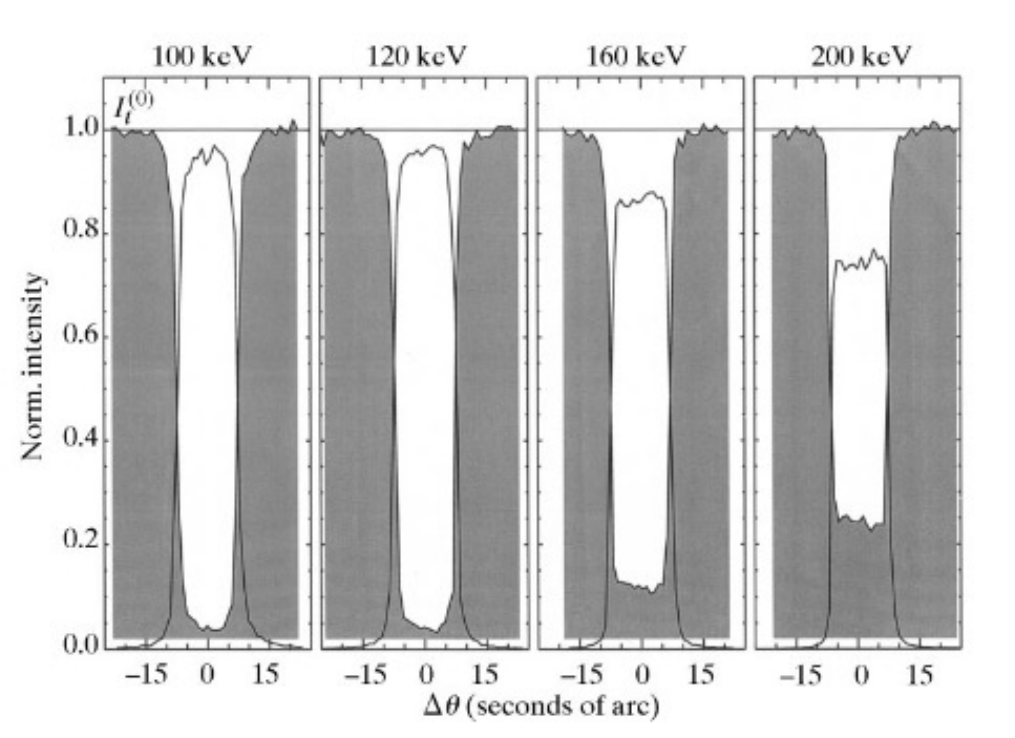}
\includegraphics[angle=0, width=0.49\textwidth]{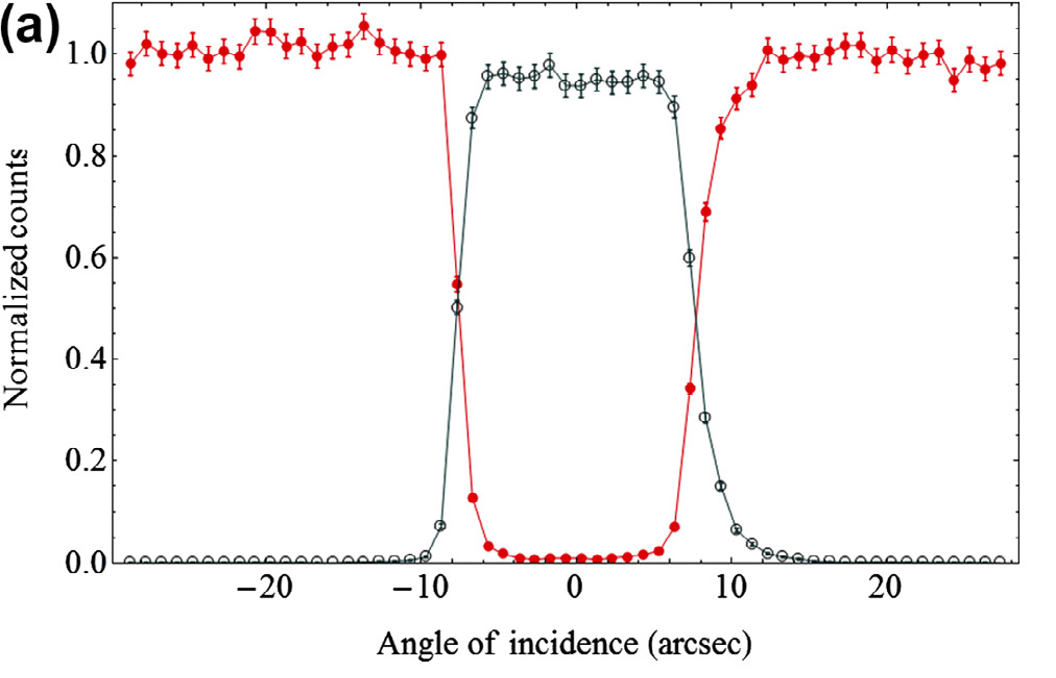}
\end{center}
%\vspace{-0.5cm}
\caption{{\em Left panel}: Diffracted and transmitted intensities as a function of the rocking angle for reflection (111) using synchrotron radiation at different energies. The transmitted intensity $I_0^{(0)}$ is normalized to 1. Reprinted from \cite{Keitel99}.
{\em Right panel}: Measured 150 keV reflectivity of a Si(111) tile 2~mm thick that was bent by performing crossed grooves in one of its surfaces. The test was performed at the European Synchrotron Radiation Facility (ESRF) in Grenoble (France). Reprinted from \cite{Guidi13}.}
\label{f:refl-bentSi(111)}
\end{figure}

\subsection{Tests of quasi-mosaicity crystals}

As discussed above, not all curved crystallographic planes generate a secondary curvature in the diffracting planes perpendicular to them. Only bending particular crystallographic planes give rise to CDP crystals. In addition, even in these cases, it cannot be excluded that a mosaic structure is produced with a consequent low crystal reflectivity  (max 50\%).
Thus experimental tests are needed to verify the crystal reflectivity to confirm the presence of a QM curvature of the diffracting planes (see eq.~\ref{e:refl_curved}).

Several experimental tests of the quasi-mosaicity production have performed. In Fig.~\ref{f:refl-bentSi(111)} (left panel) it is shown the measured reflectivity in the case of an ingot of Si$_{1-x}$Ge$_x$ crystal, with $0.01<x<0.07$, grown along the $[111]$ direction with the Czockralski technique \cite{Keitel99}, while in the same figure (right panel) it is shown  the reflectivity of Si(111) crystal tile, which was bent by grooving one of the cross sections \cite{Guidi13}. In both cases we can see very high reflectivities ($>$ 50\%) and thus the formation of a QM structure.

\section{Best crystal materials of a Laue lens}
\label{s:cryst-mat}

\begin{figure}
\begin{center}
\includegraphics[angle=0, width=0.6\textwidth]{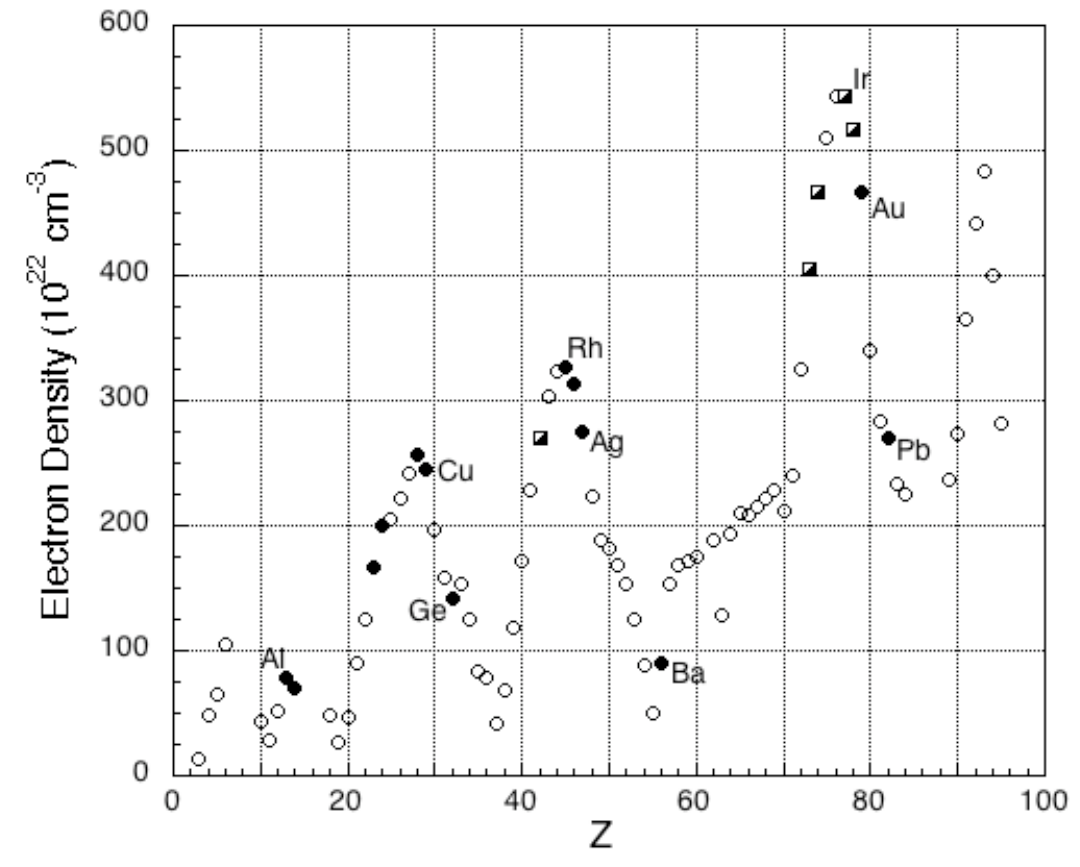}
\end{center}
%\vspace{-0.5cm}
\caption{Density $N$ of a crystal unit cell versus atomic number Z of the elements. Reprinted from Ref.~\cite{Frontera10}.}
\label{f:density.vs.Z}
\end{figure}

Independently of the crystal structure (mosaic or curved), in order to optimize the
crystal reflectivity (eq.~\ref{e:refl_mosaic} or eq.~\ref{e:refl_curved}), it is important to maximize the square ratio $|F_{hkl}/V|^2$ between the structure factor of the chosen lattice
planes $F_{hkl}$ and the volume $V$ of the unit cell. Thus it is important to minimize $V$, i.e.,  to maximize the inverse of $V$, which is the
number density $N$ of unit cells. The values of $N$ for all elements is shown in Fig.~\ref{f:density.vs.Z}.
%
% Figure 9
%
%
%
%
% Figure 10
%
\begin{figure}[ht]
\begin{center}
\includegraphics[angle=0,width=0.8\textwidth]{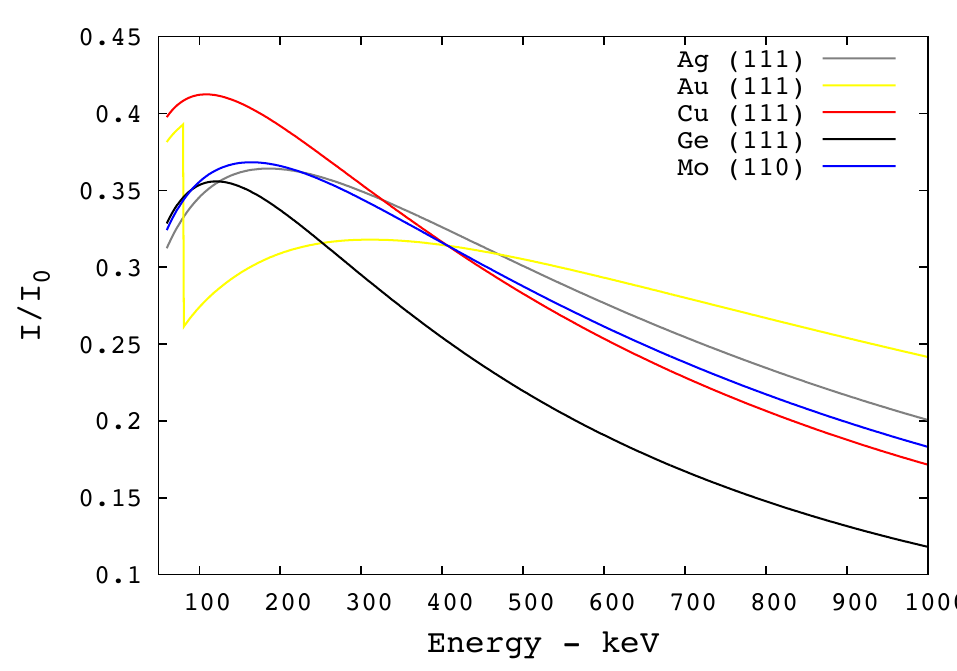}
\caption{Peak reflectivity of 5 candidate mosaic crystal materials. The Miller indices used give the highest reflectivity. A 
mosaicity of 40 arcsec is assumed. The thickness has been optimized. Adapted from \cite{Frontera10} }
\label{f:mosaic-refl-peak}
\end{center}  
\end{figure}
As can be seen, for single--element materials, broad density peaks are apparent in correspondence of the atomic numbers
5, 13, 28, 45, and 78. Common hard materials like Al ($Z=13$), Si ($Z=14$), 
Ni ($Z=28$), Cu ($Z=29$),  Zn ($Z=30$), Ge ($Z=32$), Mo ($Z=42$), Rh ($Z= 45$), Ag ($Z=47$), 
Ta ($Z=73$), W ($Z=74$), Au ($Z=42$) are good candidates to be used for Laue lenses and should 
be preferred to other elements if they are available with the requested properties.
The procurement of wafers of high quality perfect crystals of Silicon and Germanium is no more a complex issue, given the extended use of these crystals for micro-electronics (Si) and for high energy resolution X--ray detection (Ge).
Also mosaic crystals of various elements (e.g. Cu) or double-element materials (e.g., GaAs, InAs, CdTe, CaF$_2$) with the required spread can be procured. 
Peak reflectivities versus energy of few mosaic crystal materials is shown in Fig.~\ref{f:mosaic-refl-peak}.

Clearly the best lattice planes (\textit{hkl}) are those that optimize the structure factor $|F_{hkl}|$ and give rise to a quasi-mosaic structure.

\subsection{Best crystal thickness}

The crystal thickness is another crucial parameter for the reflectivity optimization of crystal. We discuss separately mosaic and CDP crystals.
%
% Figure 12
%
%\begin{figure}[ht]
%\begin{center}
%\includegraphics[angle=0,width=0.90\textwidth]{Figures/fig-thickness_vs_mos-cryst.pdf}
%\caption{The best thickness that maximize the mosaic crystal reflectivity, for various
%materials. The mosaicity assumed is 1 arcmin, the crystallite thickness
%is 1~$\mu$m, and the chosen diffracting plane is $(111)$. Adapted from \cite{Frontera10}.}
\label{f:thickness}
%\end{center}
%\end{figure}
%
%
%\subsubsection{Mosaic crystals}

In the case of mosaic crystals, from eq.~\ref{e:refl_mosaic}, it can be shown that the best crystal thickness that maximizes its reflectivity  is given by
\begin{equation}
T_{best} = \frac 1 {2 \sigma} 
\ln {\left ( 1 + \frac {2\sigma \gamma _0}{\mu} \right )} \, 
\label{eq:tmbest}
\end{equation}
From the reflectivity equation (Eq.~\ref{e:refl_mosaic}), it results that also the 
crystallite thickness $t_0$ has an important role for the reflectivity  optimization. For fixed values
of the mosaicity and crystal thickness, the highest reflectivity is obtained for a crystallite thickness
$t_0 \ll \Lambda_0$
%\label{eq:best-t_0}
%\end{equation} 
%

In general, that implies a thickness of the order of 1~$\mu$m, which is not easy to get. For example, from extended tests performed on Cu(111) supplied by ILL \cite{Courtois05}, it results that the condition above
is satisfied in single points \cite{Pellicciotta06}, but not when 
the entire crystal cross section is irradiated (values even higher than 100~$\mu$m have been found \cite{Barriere09}).
%
%The best crystal thickness for some elements is shown in Fig.~\ref{f:thickness}.
%
%\subsubsection{CDP crystals}

In the case of CDP crystals, from eq.~\ref{e:refl_curved}, it can be shown that the best crystal thickness that maximizes the reflectivity of a  CDP (or quasi-mosaic) crystal is given by
\begin{equation}
T_{best} = {\Omega_{QM} \,  \ln \left( 1 + {M \over N} \right) \over M}.
\label{e:tcbest}
\end{equation}

where $M$ and $N$ are given by
\begin{equation}
    M = \frac{\pi^2 d_{hkl}}{\Lambda_0^2} 
    \qquad
    N = \frac{\mu \Omega_{QM}}{cos \theta_B}.
\label{e:M&N}
\end{equation}

%\end{equation}
%
%while $N$ is given by
%\begin{equation}

%\label{e:N}
%\end{equation}
%
As can be seen, the best crystal thickness depends on the absorption coefficient $\mu$ and thus on photon energy $E$. 
A high absorption coefficient requires a low crystal thickness for the reflectivity maximization, i.e., a higher thickness is required at higher energies where the absorption coefficient becomes small.

\subsection{Lens sensitivity}

The sensitivity of a Laue lens can be derived as in the case of a generic focusing instrument. It is related to various lens quantities: the lens effective area $A_{eff}(\Delta E)$, which is a function of $\Delta E = E_2 - E_1$ ($E_1<E<E_2$), the area of the focal spot $A_d (\Delta E)$ in which a fraction $f_{ph} (\Delta E)$ of photons from a point-like celestial source are reflected by the lens (the total photon distribution is known as response function or Point Spread Function (PSF) of the lens), count background $B(\Delta E)$ of the focal plane position sensitive detector (PSD) and its detection efficiency $\epsilon_d (\Delta E)$.

\subsubsection{Sensitivity to continuum emission}
Assuming that, in the exposure time $T$ to a celestial source, the number of background and source counts is much higher than 1, i.e., that the statistics is Gaussian, the sensitivity at energy $E$ in the band $E_1<E<E_2$ to continuum emission at significance level of $n$ $\sigma$ ($n=$ 3 to 5), is given by
\begin{equation}
I_{C}^{min} (E) = \frac{n \sqrt{2 B A_d}}{A_{eff}(\Delta E) \, f_{ph}(\Delta E) \epsilon_d (\Delta E)  \sqrt{T} \, \sqrt{\Delta E}}
\label{e:sens}
\end{equation}
where $I_C^{min}$ (photons~cm$^{-2}$~s$^{-1}$~keV$^{-1}$) is the minimum detectable intensity  in the interval $\Delta E$  around $E$, and the background $B$ is given in counts~cm$^{-2}$~s$^{-1}$~keV$^{-1}$.

By introducing the focusing factor $G (E)$:
\begin{equation}
G(E)= f_{ph} \frac{A_{eff}(E)}{A_d(E)}
\end{equation}

it results that the lens sensitivity for continuum emission is given by:  
\begin{equation}
I_{min} (E) = \frac{n}{\eta_d \, G} \sqrt{\frac{2 B}{A_d \, T \, \Delta E}}
\label{e:sens2}
\end{equation}

If the factor $f_{ph}$ is taken to be 0.5, the area $A_d$ is called {\bf half power area}, and, given that $A_d$ is generally circular, its radius $r_d$ is called {\bf half power radius}.

\subsubsection{Sensitivity to lines}

Assuming that the line is superimposed to a continuum spectrum $I_C(E)$ (ph/cm$^2$~s~keV) and has an intrinsic Gaussian shape with mean value $E_L$ and standard deviation $\sigma_i$ (intrinsic fwhm $\Delta E_i = 2.35 \sigma_i$), it can be shown that the minimum detectable intensity of the line (ph/cm$^2$~s) is given by:

\begin{equation}
I_{min} (E_L) = \frac{1.31 \, n \, \sqrt{[2 B(E_L)\, A_d + I_C(E_L)\,f_{ph} \epsilon_d A_{eff}]\Delta E_t}}{A_{eff} \, f_{ph} \,\epsilon_d \,  \sqrt{T}}
\label{e:line-sens}
\end{equation}
with most of the terms already defined above, 
$\Delta E_t = \sqrt{\Delta E_i^2 + \Delta E_d^2}$, where $\Delta E_d$ is the fwhm of the photon spread of the line due to the detector energy resolution, and the constant 1.31 is due to the assumption that we count only the line photons within the total fwhm $\Delta E_t$.  

\subsection{Broad vs. narrow band Laue lenses} 

Two lens classes can be identified, \emph{broad band} Laue lenses and \emph{narrow band} Laue lenses, the former to cover a broad energy band (e.g., 50-600 keV)
for the study of also continuum source spectra, the latter to achieve an optimal sensitivity 
in a relatively narrow energy band (e.g., 450-550 keV) for gamma-ray line spectroscopy studies.

These two classes of lenses require different criteria in the crystal material and diffracting planes to be used for its optimization.

\subsubsection{Narrow-band Laue lenses}

Assuming the usual ring--like geometry, in order that all the crystals reflect photons in a given narrow energy band, centred, e.g. at energy $E_0$, from eq.~\ref{e:lens-energy}, different crystalline planes \emph{(hkl)} are needed for  each ring.
Indeed, for a given focal distance of the lens, crystal rings with a~radius
$\mathrm{r}_{2} > \mathrm{r}_{1}$, require spacing between crystal planes $d_2 < d_1$ , such that $d_{hkl} \times r = const$ to concentrate incident photons in the the same energy passband. This can be obtained by changing the crystal material or the diffracting planes of  the same material (e.g., higher order diffraction).  From eq.~\ref{e:dhkl-and-V}, for materials with a~cubic structure (e.g., the face-centered cubic cell of copper, germanium or silicon), for which $d_{hkl}$ is inversely proportional to the sum $\sqrt{\mathrm{h}^2+\mathrm{k}^2+\mathrm{l}^2}$, the ring radii are proportional to this quantity. 

In the case the higher order diffraction criterion is adopted, the diffraction efficiency decreases, but the outer ring, in which the crystal with a lower $d$ has to be located, has a larger area, and can compensate the lower efficiency. Thus all rings could contribute by about the same amount to the effective area in the required narrow band. 

An example of a~narrow bandpass Laue lens was the balloon experiment CLAIRE \cite{vonBallmoos05} that I will discussed later.

A lens made of mosaic Copper crystals designed to focus soft gamma-rays in the band from 100 to 200 keV with centroid at 140.6 keV, was assembled at Argonne National Laboratory, for radiotherapy applications \cite{Roa01}.
\subsubsection{Broad-band Laue lenses}

Broad-band Laue lenses are the best solution for getting a gamma--ray focusing telescope. These telescopes can be exploited for the deep studies of continuum and line emission spectra of celestial gamma-ray sources. In order to get the best focusing and the highest effective areas of these lenses, curved crystals with quasi-mosaic structure of the diffracting planes and with primary curvature radius equal to the curvature radius of the lens, are the best solution. 

Assuming a ring-like geometry of the lens, for a given diffracting plane of the adopted crystal (e.g., Si(111)),  contiguous rings focus
slightly different photon energies (see eq.~\ref{e:lens-energy}), and thus several rings can cover a broad energy band. By using different bent crystal materials (e.g., Si(111) and Ge(111)), a broad band can be efficiently covered. The effective area of the lens is due not only to the first order diffraction, but also to higher orders. Higher order diffraction can also be exploited to extend the lens passband.

A~gamma-ray lens with a broad passband (300 keV--1.5 MeV) was proposed in the 90's by N.~Lund \cite{Lund92}. He assumed mosaic crystals of Copper and Gold. In order to achieve a significant effective area
at high energies ($350\,\mathrm{cm}^2$ at 300\,keV and $25\,\mathrm{cm}^2$ at 1.3\,MeV), the focal length requested was 50 m.

A broad passband (200-1300 keV) Laue lens telescope was also proposed for the Gamma-Ray Imager (GRI) formation flying mission concept, submitted for the ESAs Cosmic Vision 2015–2025 program. The lens was made mainly of mosaic crystals of Copper, with a focal length of 100 m \cite{Knodleseder09}.

A first study of a broad band lens (90--600 keV) with a distribution of the bent crystals on various contiguous rings and a 20 m focal length, was performed by us \cite{Virgilli17}.

\section{Optical properties of a lens}
\label{s:optical-properties}

Independently of the the reflection efficiency, the optical properties of a lens have been investigated by means of Monte Carlo (MC) simulations for  mosaic crystals \cite{Pisa05b,Frontera06} and for CDP crystals (see, e.g., \cite{Virgilli17}).

\subsection{Lens with mosaic crystals}

\begin{figure}
\begin{center}
\includegraphics[angle=0, width=0.49\textwidth]{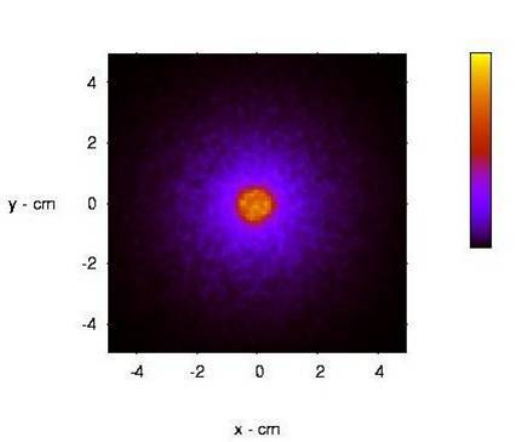}
\includegraphics[angle=0, width=0.49\textwidth]{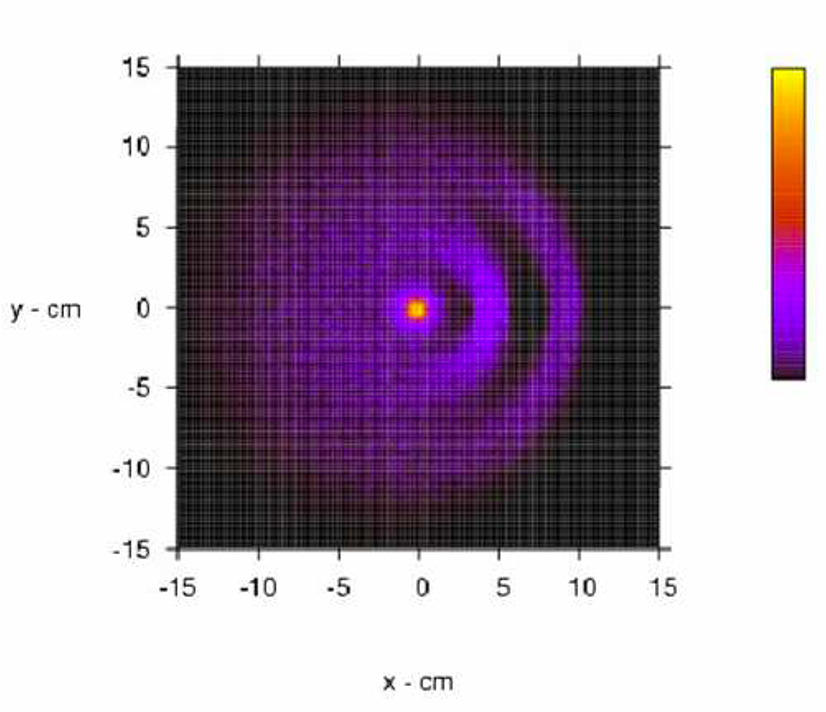}
\end{center}
\vspace{-0.5cm}
\caption{Response function (PSF) of a lens made of mosaic crystals with 40 m focal length and 150--600 keV passband. {\em Left panel:} On axis PSF. {\em Right panel:} lens response funtion in the case of three sources, two of them at 2 and 4 arcmin off-axis. Reprinted from 
\cite{Frontera08a}.}
\label{f:psf_mosaic}
\end{figure}

In the case of mosaic crystals, the derived Point Spread Function (PSF) depends on the crystal size, on their mosaicity and on the accuracy of their positioning in the lens. 
Studies of the PSF with mosaic crystals have been performed by different authors (see, e.g., \cite{Pisa05b,Barriere09c}. In Fig.~\ref{f:psf_mosaic} (top panel) it is shown the on--axis PSF for a ring--shaped lens of 40 m focal length, 150--600 keV energy passband and a crystal tile cross section of 10$\times$10~mm$^2$ \cite{Frontera08a}. In this case it is assumed that the crystals have 1 arcmin mosaic spread and that they are properly oriented in the lens. 

In the same figure (bottom panel) it is shown, for the same lens configuration \cite{Frontera08a}, the expected PSF when three sources are in its Field of View (FOV).  In  this case one of the sources is on--axis and the other two off--axis. As can be seen, due to coma aberration, in the case of the off--axis sources the image has a ring shape centred in the lens focus, with the radius that increases  with the offset angle, 
and with a non uniform distribution of the reflected photons with azimuth. It is found also that the integrated number of photons focused by the  lens does not significantly vary from an on--axis  to off-axis source, but the space distribution, for off-axis sources, spread over an increasing surface. As a consequence, the FOV of the lens is determined by the detector radius, with a lens sensitivity that decreases with source offset.
The azimuthal non-uniformity of the PSF for off-axis sources could be exploited for deriving information on the azimuthal source direction.

For a lens made of mosaic crystals, its angular resolution depends on the mosaic spread and on the misalignment of the lens crystal tiles in the lens. For the lens image shown in Fig.~\ref{f:psf_mosaic}, the angular resolution is of the order of 1 arcmin.  

%
% Figure  11
%

%
% Figure  12
%
\begin{figure}[ht]
\begin{center}
\includegraphics[angle=0, width=0.6\textwidth]{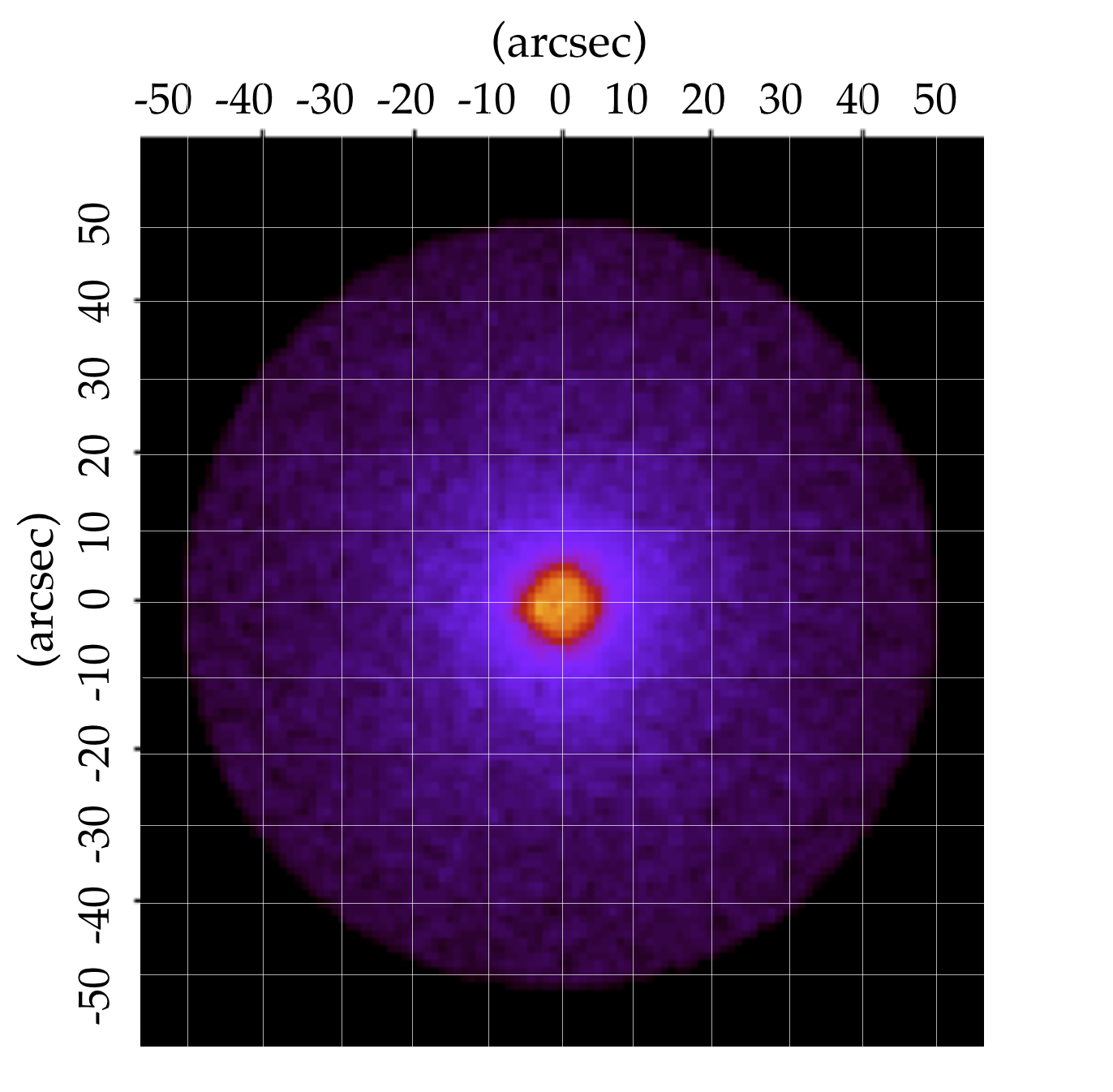}
\includegraphics[angle=0, width=0.8\textwidth]{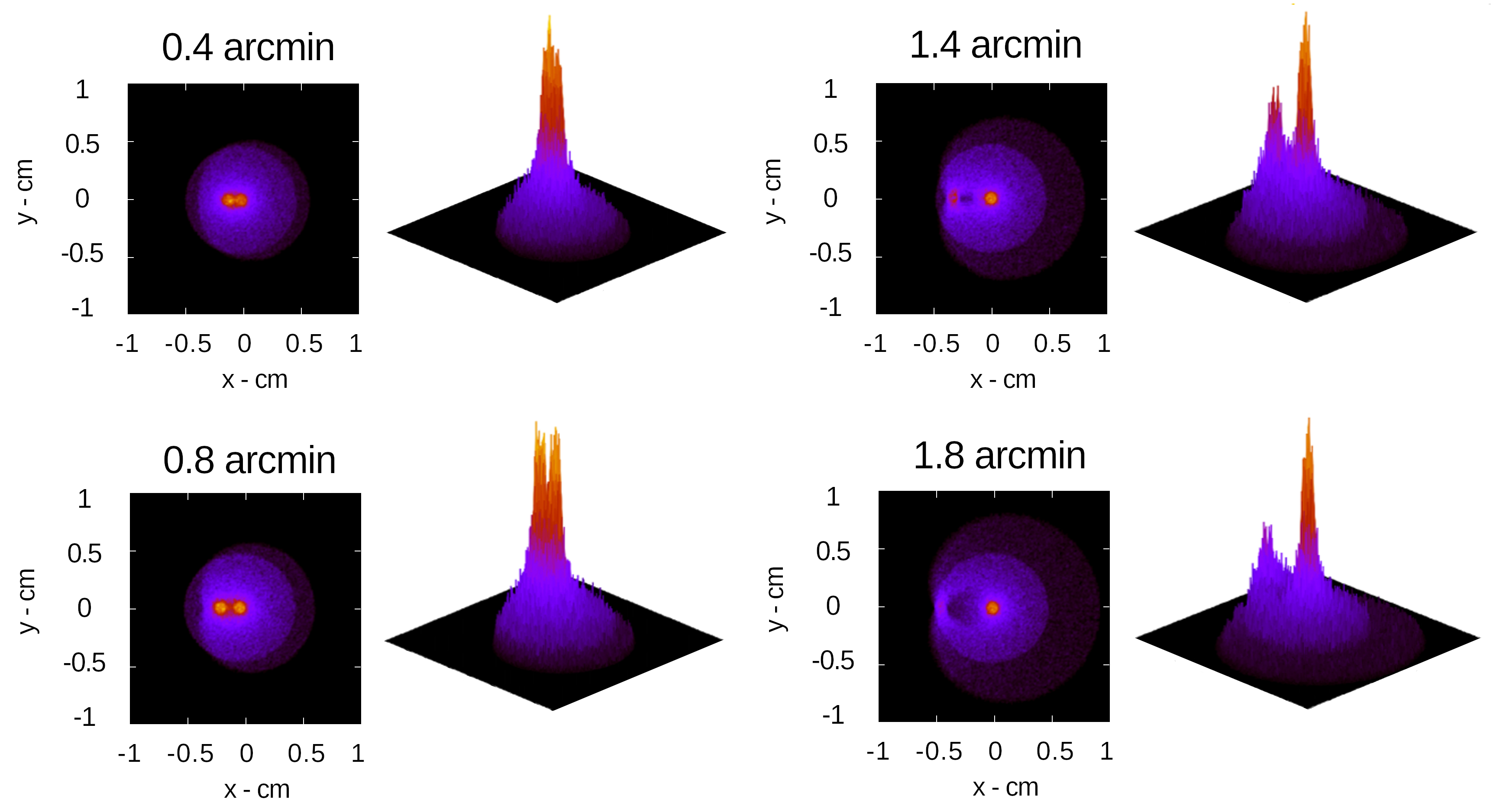}
\end{center}
%\vspace{-0.5cm propos}
\caption{Expected response function (PSF) of a 20 m focal length Laue lens telescope, proposed for the \astena\ mission concept (see, e.g., \cite{Frontera21}), made of bent crystal tiles of Si(111) and Ge(111) with curvature radius of 40 m, CDP diffracting planes, and 50-600 keV passband. {\em Top panel}: on-axis PSF. The half-power diameter is 1.5 mm. {\em Bottom panel}: off-axis PSF for increasing values of the offset angle of the sources.
Reprinted from \cite{Frontera21}.}
\label{f:psf_bent-crystals}
\end{figure}

\subsection{Lens with bent quasi-mosaic crystals}
\label{s:optical-CDP}

As discussed in section \ref{crystal-focusing}, the best focusing is obtained in the case of bent crystals with curvature radius equal to that of the spherical lens.
In Fig.~\ref{f:psf_bent-crystals}, top panel, it is shown the on-axis PSF of a Laue lens made of bent crystals of Si(111) and Ge(111) with quasi-mosaic diffracting planes, a passband from 50 to 600 keV and 20 m focal length \cite{Frontera21}, proposed for the \astena\ mission concept we will discuss later (see section~\ref{s:Concluding}). The expected angular resolution is 30 arcsec for a positioning accuracy of the crystal tiles $<$10 arcsec.

The off-axis PSF of the same lens is shown in the bottom panel of Fig.~\ref{f:psf_bent-crystals}. Such PSF is clearly better than that obtained in the case of mosaic crystals, even if it is clear that for off-axis sources the PSF is becoming worse due to coma aberration. The FOV of the lens is estimated to be about 4 arcmin.

\section{Focal plane detectors for Laue lenses} 
\label{s:focalPSD}

A Laue lens telescope, like all the focusing telescopes, requires a position sensitive detector (PSD) in its focal plane. In the case of Laue lenses with a passband in the soft gamma-ray astronomy, in order to face the astrophysical issues still open in this band, some of which discussed in section~\ref{s:intro}, it is crucial to have, in addition to a position sensitivity also an excellent energy resolution of the focal plane detector.
Several studies of PSDs for soft gamma-ray astronomy and thus also for Laue lens telescopes, have also been performed for both Cadmium Zinc Telluride (CZT, see, e.g., \cite{Kuvvetli10,Kuvvetli14,Caroli22} and references therein) and for High Purity Germanium  detectors (HPGe, see, e.g., \cite{Boggs23} and references therein). The advantage of CZT is that it can be operated at environment temperature at expenses of a lower energy resolution than the HPGe, that, however, requires to be operated at the liquid Nitrogen temperature. In the latter case, for satellite mission, this temperature can be achieved using Stirling motors.
A possible configuration of a CZT PSD for the Laue lens telescope proposed for the \astena\ satellite mission concept is discussed in Ref.~\cite{Frontera21}.
A 3-dimensional (3D) PSD in the focal plane of a Laue lens can also be used Compton polarimeter, that can achieve even unprecedented sensitivity (see, e.g., \cite{Caroli18}). 

%A possible configuration of a focal plane PSD based on CZT for the ASTENA mission concept (see section~\ref{s:Concluding}) is described in \cite{Virgilli22} and it is shown in Fig.~\ref{f:CZT-focal-plane}.
%
% Figure  15
%
%\begin{figure}[ht]
%\begin{center}
%\includegraphics[angle=0, width=0.5\textwidth]{Figures/fig-Astena-CZT-PSD1.pdf}
%\includegraphics[angle=0, width=0.4\textwidth]{Figures/fig-Astena-CZT-PSD2.png}
%\end{center}
%\vspace{-0.5cm}
%\caption{A possible focal plane PSD, based on CZT, proposed for the ASTENA mission concept. The detector is made of a set of detection modules. {\em Left panel}: a view of the entire detector. {\em Right panel}: a view of a single module.
%Reprinted from \cite{Virgilli22}.}
%\label{f:CZT-focal-plane}
%\end{figure}

\section{Laue lens development constraints}

Two key issues have to be faced in order build a Laue lens: 
\begin{itemize}
\item Development of technologies for the massive production  of crystals with the required properties;
\item Development of technologies for assembling, in a reasonable time, consistent with that of preparation of a space mission, thousands of crystal tiles in a lens with the proper orientation accuracy. This issue is the most critical one. It crucially depends on the focal length of the lens. 
Higher focal lengths require higher positioning accuracy, thus the development is more challenging for lenses working at the highest energies. 
\end{itemize}
The technologies adopted for building Laue lenses have been changed with time, depending on the goals that would be achieved  and the available technologies  that could be exploited.

\section{First generation of Laue lenses}

\subsection{The first developed Laue lens}

The first developed Laue lens, with a broad energy band (20--140 keV), was made of rock salt crystals with mosaic structure \cite{Lindquist68}. It was mounted on a paraboloidal frame of about 180 cm diameter with a focal length of about 290 cm. The lens telescope, with a focal plane detector of NaI(Tl) of 5~cm diameter, was flown aboard a stratospheric balloon. In spite of the large area of the lens, the estimated effective area was very small: about 160~cm$^2$ in the 20--40 keV band. Instead the angular resolution (FWHM) was very good for the epoch in which the lens was launched: 1 degree.

After this earliest experiment, in the last two decades Laue lens developments have been being carried out in different institutions. We summarize here the major results obtained.

\subsection{The CLAIRE narrow band Laue lens}

The CLAIRE Laue lens~\cite{Ballmoos05} was developed for a balloon experiment. It was a narrow band lens with an energy centroid of 170 keV and a bandwidth of 3 keV. Goal of the balloon experiment was to demonstrate that
a Laue lens prototype can work under space conditions, measuring its performance by observing an astrophysical target. The experiment was flown twice (2000, 2001) on a stratospheric balloon. The lens was also tested on a 205 m long optical bench~\cite{alvarez04}.

%\subsubsection{Lens crystals for CLAIRE and their production}
The lens, with a focal length of 279 cm, consisted of 556 crystals of Ge$_{1-x}$Si$_x$ ($x\sim 0.02$), with a mosaic structure from 30 arcsec to 2 arcmin and a cross section of $10\times10$~cm$^2$. The crystals were grown and cut at the Institut f\"ur Kristallz\"uchtung (IKZ) in Berlin, by a modified Czochralski technique~\cite{Abrosimov05a,Abrosimov05b}.

The crystals were mounted on eight rings on a Titanium frame of 40 cm diameter. In order to assemble the crystals in the lens, the individual tiles were mounted on flexible aluminium supports, which in turn were mounted on the lens frame. 
The tuning of the crystals was performed by mechanically tilting each crystal tile to the appropriate Bragg angle so that the diffracting energy was 170 keV for a source at infinity. 
%A view of the assembled lens is shown in Fig.~\ref{f:Claire}.

%
% Figure  15
%
%\begin{figure}[ht]
%\begin{center}
%\includegraphics[angle=0, width=0.8\textwidth]%{Figures/fig-Claire-lens.pdf}
%\end{center}
%\vspace{-0.5cm}
%\caption{Claire Laue lens during its assembling and crystal tuning.
%Reprinted from \cite{Ballmoos05}.}
%\label{f:Claire}
%\end{figure}

%The geometric area of the CLAIRE lens was
%$511\,\mathrm{cm}^2$ with a 
The measured peak efficiency from ground test was of the order of 10\%, while the FOV was 1.5 arcmin. 
The photons were focused onto a focal plane detector made an array $3\times3$~array of HPGe modules housed in a~single cylindrical aluminum cryostat. For more details on the experiment and Crab observation results see \cite{Ballmoos05}.

\subsection{Laue lenses with tunable focal distance} 

Laue lenses with tunable focal distance have been proposed to achieve different goals. 

 The  first tunable $\gamma$-ray lens prototype was developed and demonstrated by a collaboration project between Argonne National Laboratory (ANL) and the French space agency CNES \cite{Kohnle98}.  The goal was that of observing more than one astrophysical line with a narrow band Laue lens. This goal requires the tuning of the Bragg angle and focal distance. 
The strategy adopted to change the Bragg angle was that of using piezo-driven actuators and an eddy-current sensor to determine the current position. In this way, an accuracy of 0.1$-$0.4\,arcsec could be achieved. The drawback of this strategy is the large volume required, the decreased filling factor of crystal tiles, and the additional thickness that the reflected photons had to cross, in addition to the required power and the additional mass of the lens.

Another concept of a Laue lens with tunable focal distance has been recently suggested \cite{Lund21a,Lund21b}. The goal is to cover, with a single lens, a range of energies from 200 keV to 2.5 MeV, using mosaic crystals of Silver and Copper, and a range of focal distances from 50~m to 400~m. Obviously, in this case, two satellites in formation flying, are needed: one with the lens and the other with the focal plane detector.  The proposed technology to orient each crystal is the design of a light-weight crystal adjustment pedestal and an optical system for verifying the inclinations of all crystals in the lens. Details of the technology are reported in \cite{Lund21b}.

\subsection{HAXTEL broad band Laue lens}

The science goal of the HAXTEL (HArd X-ray TELescope) was that of building a broad band (70/100-600 keV) Laue lens.
The crystals adopted were mosaic crystals of Cu(111) developed and produced at ILL \cite{Courtois05}. The tile cross-section was 15$\times$15 mm$^2$ while its thickness was 2 mm. The mosaic spread  of the crystals ranged from $\sim 2.5$ and $\sim 3.5$ arcmin \cite{Frontera08a}.

%
% Figure 16
%
%\begin{figure}
%\begin{center}
%\includegraphics[angle=0, width=0.6\textwidth]{Figures/fig-LARIXA-view.jpg}
%\includegraphics[angle=0, width=0.4\textwidth]{facility_view1.pdf}
%\end{center}
%\vspace{-0.5cm}
%\caption{A view of the current configuration of the hard X-ray facility LARIX A of the University of Ferrara.}
%\label{f:facility}
%\end{figure}
%
For HAXTEL, it was developed a technology that did not require any mechanism for a fine adjustment of the crystal orientation once the crystal was positioned in the lens frame.
This technique, later patented, is described in \cite{Frontera07,Frontera08a}. 
%It made use of a counter-mask provided with holes, two for each crystal tile. 
%Each tile was positioned on the countermask by means of two cylindrical pins, rigidly glued to the crystal tile, that were inserted in the countermask holes. The pin direction and the axis of the average lattice plane of each crystal tile 
%had to be exactly orthogonal. The hole axis direction constrained the energy of the photons diffracted by the tile, while the relative position of the two holes in the countermask established the azimuthal orientation of the axis 
%of the crystal lattice plane. This axis had to cross the lens axis.

%Depending on the direction of the hole 
%axes in the countermask, the desired geometry of lens could be obtained. In the case of a lens for space astronomy, the hole axes had to be directed toward the center of curvature of the lens. 
%Once all the crystal tiles were placed on the counter-mask, a
%frame was glued to the entire set of the crystals. Then the lens frame, along with the crystals, was separated from the counter-mask and from the pins.

A first lens prototype, with 6 m focal length, made of a ring of 36 ~cm diameter with 26 crystal tiles, was developed and tested \cite{Frontera08a}. In this case, the selected diffracting planes were set parallel to the lens axis for a quick test with a highly divergent polychromatic  X-ray beam. The lens frame was made of carbon fibers with a total thickness of  1 mm.

\begin{figure}
\begin{center}
\includegraphics[angle=0, width=0.5\textwidth]{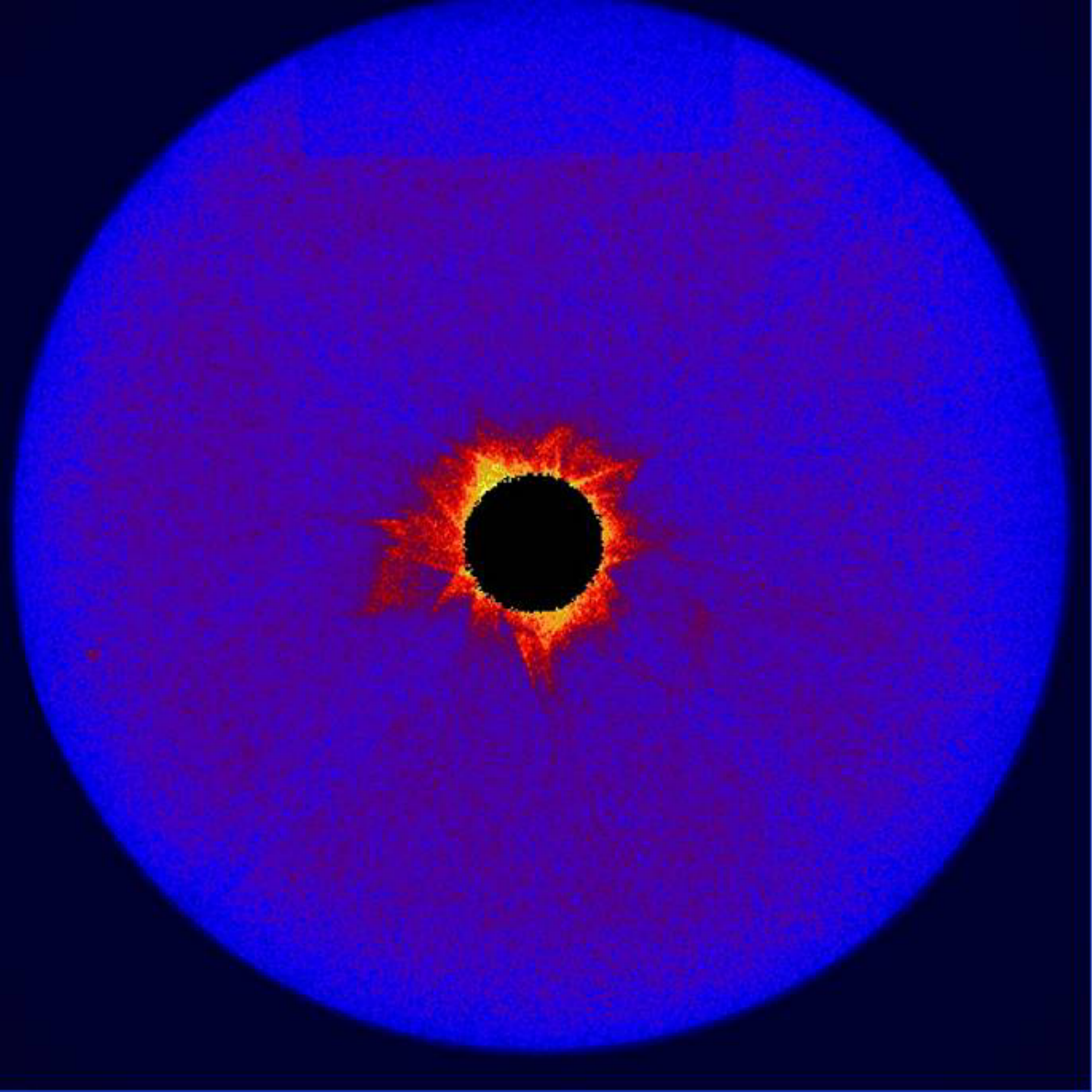}
\end{center}
%\vspace{-0.5cm}
\caption{Difference between the PSF measured
and that obtained with a Monte Carlo code by assuming a perfect positioning 
of the crystal tiles in the lens. Reprinted from \cite{Frontera08a} }
\label{f:diff_image_spread}
\end{figure}

The X--ray beam was that of the the LARIX (LArge Italian hard X--ray) facility of the University of Ferrara~\cite{Loffredo05}. The facility is now called LARIX A after its upgrade and with transnational access (\url{https://larixfacility.unife.it/}).
%is shown in Fig.~\ref{f:facility}. 
Figure~\ref{f:diff_image_spread} shows the difference between the measured PSF and that obtained from a Monte Carlo simulation, in which a perfect positioning of the crystals in the lens was assumed \cite{Frontera08b}.
%As can be seen, only the center part of the measured image (black region) is  subtracted by the simulated image. The corona
%in the difference image is the result of the error budget made during the lens assembly process.
%
% Figure 13
%

It was found that the PSF radius at which the expected fraction of focused photons reach the saturation (16 mm) corresponds to $\sim$60\% of the measured fraction. 
A better crystal orientation was later undertaken as discussed in \cite{Ferrari09}.

\section{Satellite mission concepts with the first generation of Laue lens telescopes}

Laue lens telescopes were proposed for two satellite mission concepts submitted  to ESA  in the first decade of 2000s: {\em Gamma-Ray Immager} (GRI)  and {\em DUAL}. 

The {\em GRI} mission concept \cite{Knodleseder09},   devoted to perform studies of particle acceleration processes and explosion physics, included a broad band Laue lens (200--1300 keV) with a focal length of 100~m, made of Cu, SiGe and Ge crystals with mosaic structure (mosaicity between 30 and 40 arcsec). The angular resolution was 30 arcsec, while its FOV was 6 arcmin. 

The {\em DUAL} mission concept \cite{vonBallmoos12} 
included a narrow band Laue lens (800--900 keV) with a focal length of 68~m, with the main goal of studying the 847 line emitted from $^{56}$Co to determine the
amount of $^{56}$Ni synthesized during SN explosions. 
The lens was made of Gold and Copper crystals with mosaic structure (30 arcsec mosaicity), a total geometrical area of about 6000~cm$^2$ and a total effective area of about 500~cm$^2$. 

Both {\em GRI} and {\em DUAL} were designed as formation-flying missions with two spacecraft: one carrying the lens, and the other carrying the detector.

\section{More recent projects of Laue lenses}

%
% Figure 14
%
\begin{figure}[ht]
\begin{center}
\includegraphics[angle=0, width=0.7\textwidth]{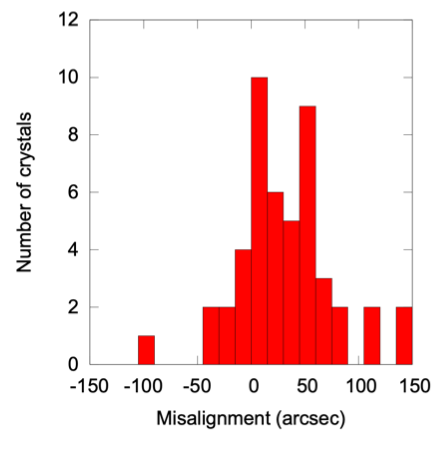}
\end{center}
%\vspace{-0.5cm}
\caption{Misalignment distribution of 48 crystal tiles used to build a Laue lens prototype, developed at University of California Berkeley. Reprinted from \cite{Wade15}.}
\label{f:misalignment-Wade15}
\end{figure}

\subsection{Laue lens development at UC Berkeley}
\label{s:Berkeley}

Goal of the project was to build a demonstrator of Laue lens, reflecting in the energy band from $\sim 90$ to $\sim 130$~keV, with mosaic crystals of Al(111), Fe(110) and Fe(200) \cite{Wade15}. For that, a dedicated X--ray beamline was developed (see details in \cite{Barriere14b} and references therein). With the X--ray beam line the crystals were correctly oriented and mounted on the substrate. However, in spite that the positioning of the crystals within a few arcsec, given that the crystal diffracting planes were misaligned ($\pm 1^o$) with respect to the cross sections, a wedge of glue was needed and thus a misalignment, due to shrinkage of the glue while curing, was found.
The reported crystal misalignment distribution of the developed prototype is shown in Fig.~\ref{f:misalignment-Wade15}.

\subsection{The LAUE project}
\label{s:LAUE}

The LAUE project was devoted to the development of high reflectivity crystals and of an advanced technology for an accurate assembling of a broad band (80--600 keV) Laue lens with a focal length of 20 m. 

%
% Figure 15
%
\begin{figure}
\begin{center}
\includegraphics[angle=0, width=0.95\textwidth]{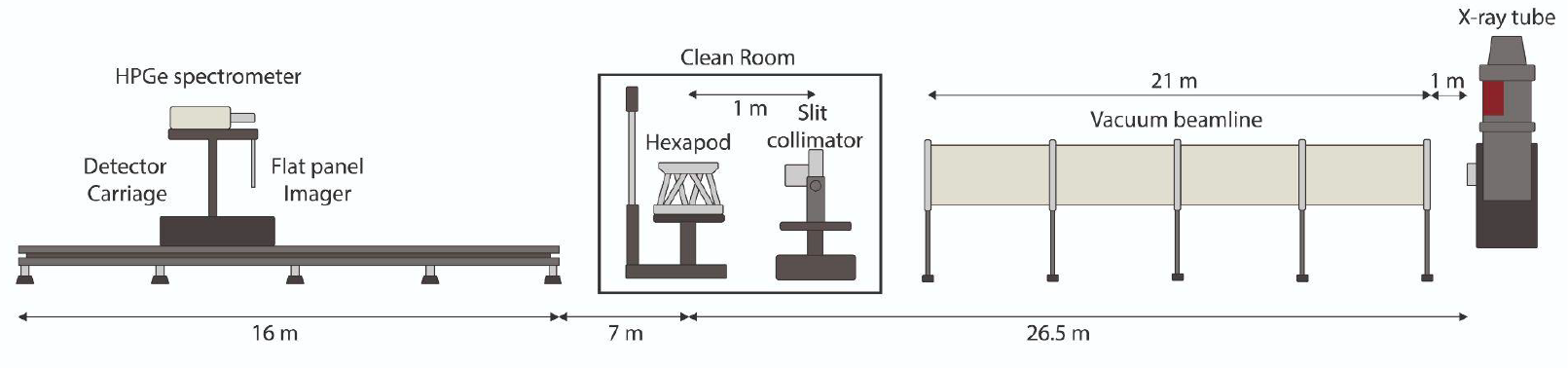}
\end{center}
%\vspace{-0.5cm}
\caption{A schematic of the LARIX~T beamline.}
\label{f:LARIX-T}
\end{figure}

Concerning crystals, we concentrated our efforts to the development of bent crystals with the same curvature of the lens (40~m) and with diffracting planes such that to get  CDP crystals.  Two technologies, already discussed in the section~\ref{s:techniques-for-CDP}, were developed. One consisted in the grooving one of the surfaces of  a Si(111) perfect crystal 2 mm thick (see, e.g.,\cite{Bellucci11}), with very high reflectivity results previously shown (see Fig.~\ref{f:refl-bentSi(111)}). The cons of this technology was the fragility of the crystal tiles due to the grooves. 
The other technology developed (lapping of one of the two surfaces of a crystal, \cite{Ferrari13}) did not give rise to a crystal fragility, but the lapping can erode the crystal surface and give rise to a mosaic structure of the diffracting planes, with the consequence of a  low reflectivity \cite{Ferrari13b}. The crystals bent with this technology were Ge(220), Si(220) and GaAs(220).

Concerning the assembly technology, for getting an accurate orientation of all crystal tiles in the lens, we bonded the crystals to their substrate under control of a parallel gamma--ray beam and using an adhesive with a very low linear shrinkage factor (0.03\%).

To get a parallel gamma--ray beam we exploited the 100 m tunnel of the LARIX laboratory of the University of Ferrara, by installing  there a long X--ray facility (LARIX T, see Fig.~\ref{f:LARIX-T}, with all the main components X--ray source collimator, exapode, detector) controlled from a console room.

%
% Figure 16
%
\begin{figure}[ht]
\begin{center}
\includegraphics[angle=0, width=0.95\textwidth]{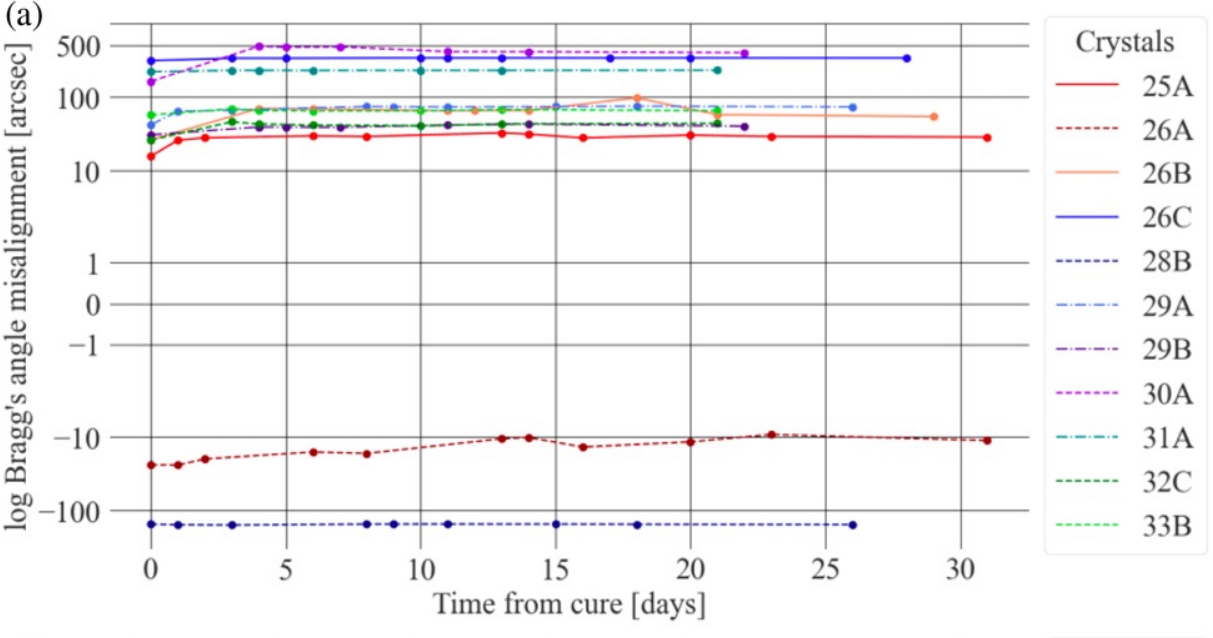}
\end{center}
%\vspace{-0.5cm}
\caption{Misalignment distribution in Bragg angle of 11 bent crystal tiles (40~m curvature radius) of Ge(220) used to build a sector of Laue lens at the University of Ferrara. Adapted from \cite{Ferro24}.}
\label{f:misalignment-Ferro24}
\end{figure}

Final results of this activity have been recently reported \cite{Ferro24}. As found in other developments (see, e.g., section~\ref{s:Berkeley}), the shrinkage of the glue while curing determined, with respect to the nominal orientation, a strong misalignment in the radial direction, and thus in the Bragg angle (see Fig.~\ref{f:misalignment-Ferro24}). Instead in the azimuth direction, i.e., at 90 deg from the radial direction, the misalignment produced by the glue is in the arcsecond range, as can be seen in the bottom panel of Fig. 8 of the paper by \cite{Ferro24}.

\subsection{Laue lenses  based on Silicon Pore Optics technology}

\begin{figure}[ht]
\begin{center}
\includegraphics[angle=0, width=0.7\textwidth]{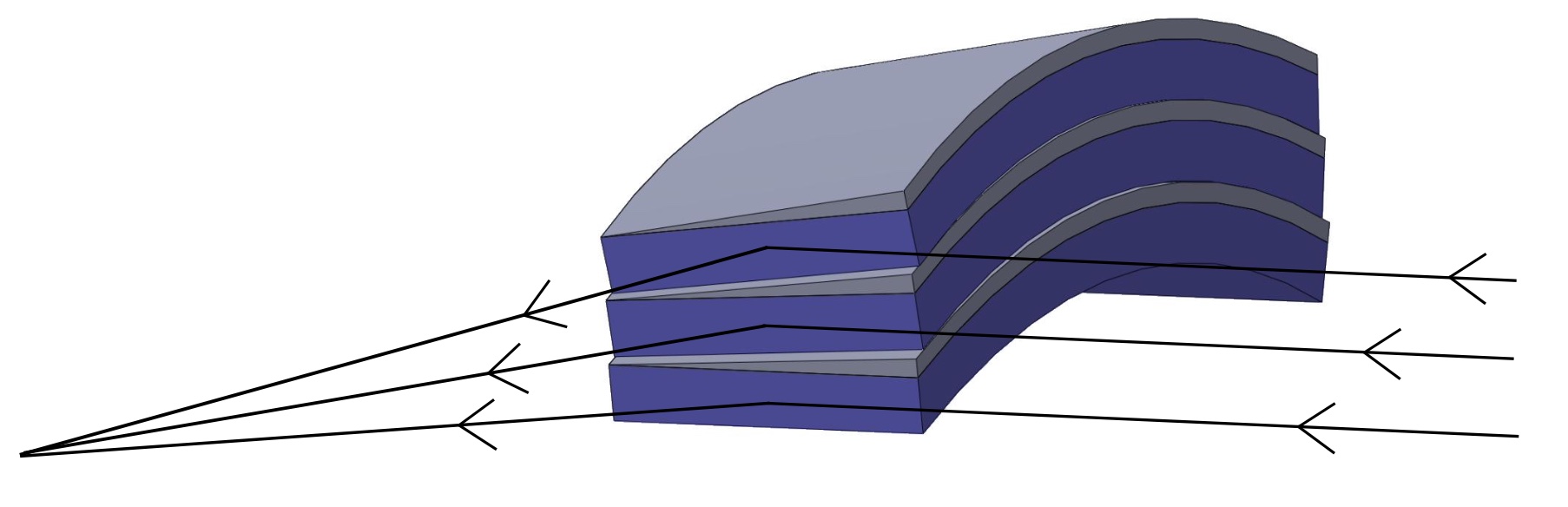}
\end{center}
%\vspace{-0.5cm}
\caption{Principle of the Silicon crystal geometry designed to be used as SiLC. Reprinted from \cite{Girou17}.}
\label{f:SiLC-Girou17}
\end{figure}

The development of the Silicon pore optics (SPO) technology for the ATHENA satellite mission concept \cite{Collon16}, has suggested other applications of the Silicon crystals, in particular to use bent and bonded thin Si crystals for building a Laue lens with 80--300~keV passband. The resulting single elements of this lens have been called Silicon Laue Components (SiLC). An example of SiLC geometry design is shown in Fig.~\ref{f:SiLC-Girou17}.

%
% Figure 17
%

The Silicon crystal tiles are assumed to be been obtained by dicing commercial double-sided polished high purity Si wafers.
Each of the tiles, with diffracting planes (111), is designed to be bent in azimuthal direction in order to fit the ring radius of the lens, while the Bragg direction (along the incident photon direction) is adjusted by inserting a proper wedge between contiguous crystal tiles in addition to an achievable radial bending of the crystal tiles. The crystal bending is expect to give a high diffraction efficiency.
The ray tracing of a lens based on SiLCs was performed \cite{Girou17} with very good results in terms of focusing area ($<1$~mm$^2$), while the study of such a Laue lens concept with a 20--158 keV passband and a focal of 8 m has been recently reported \cite{Barriere23} for polarimetry observations of celestial sources from a stratospheric balloon.   

It is crucial to perform laboratory tests to verify these promising expectations from the SiLCs for covering the hard X--ray band up to about 200~keV.

% Figure 18
%
\begin{figure}[ht]
\begin{center}
\includegraphics[angle=0, width=0.90\textwidth]{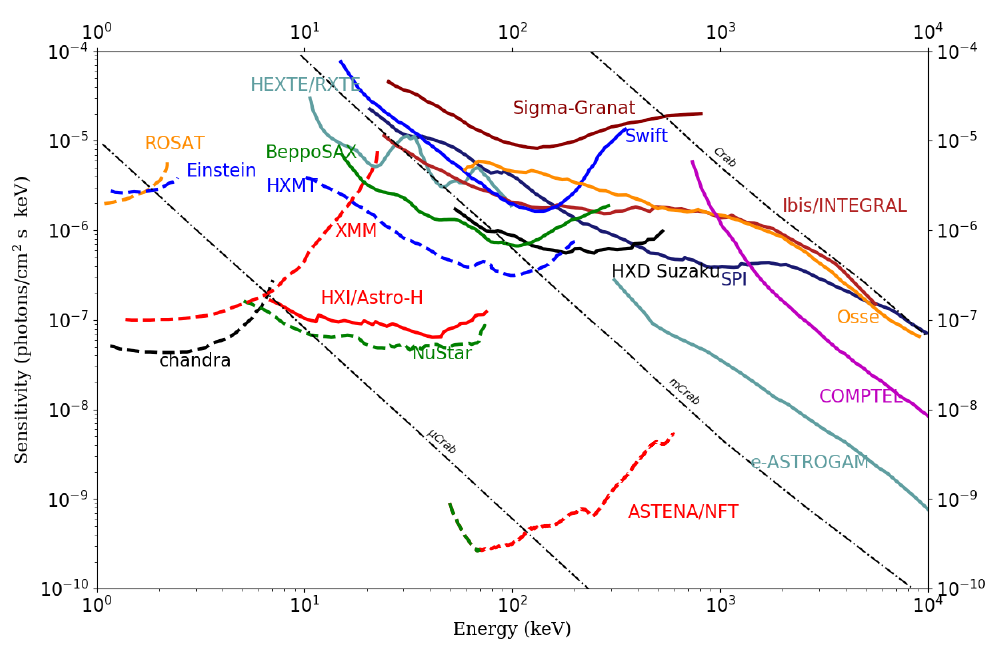}
\\
\includegraphics[angle=0, width=0.45\textwidth]{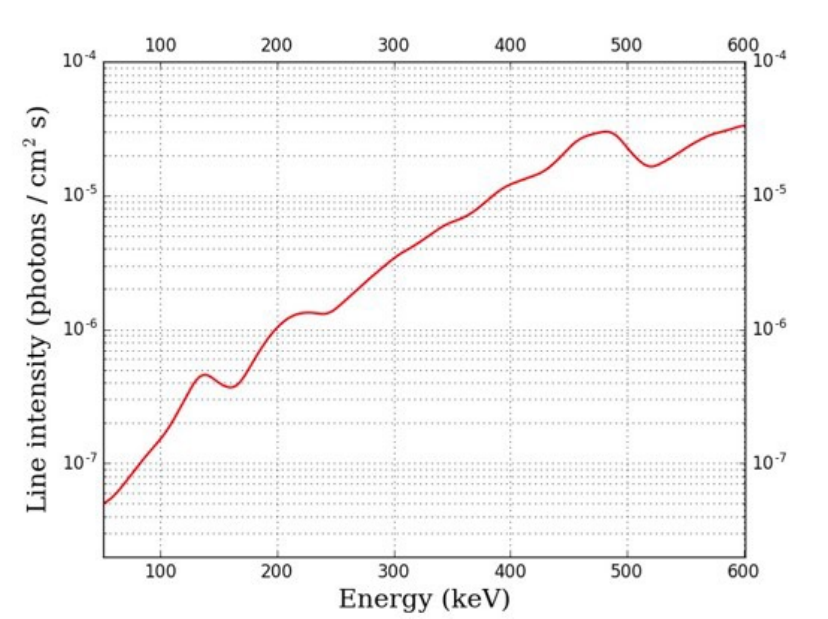}
\includegraphics[angle=0, width=0.45\textwidth]{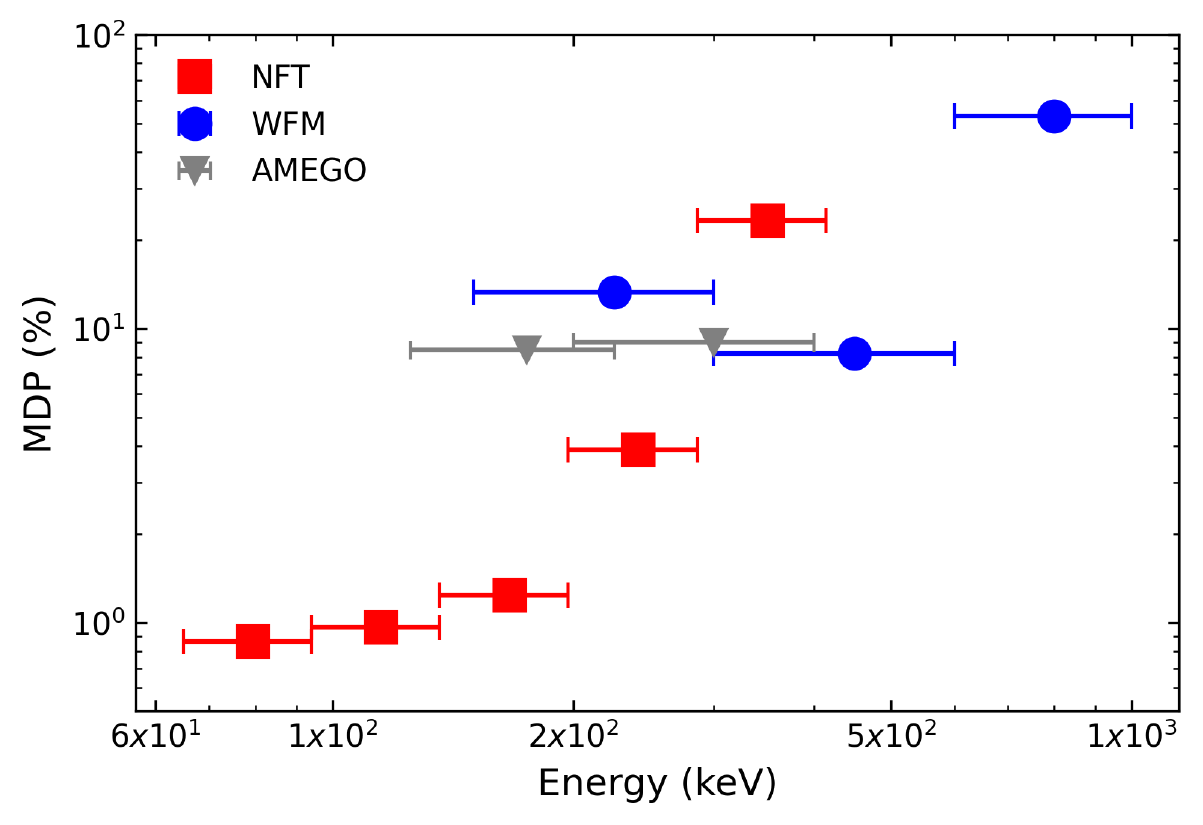}
\end{center}
%\vspace{-0.5cm}
\caption{Expected performance of the \astena\ mission concept. {\em Top left panel}: NFT sensitivity to continuum spectra at 3$\sigma$ confidence level, with bandwidth $\Delta E = E/2$ and observation time $T = 10^5$~s. {Top right panel}: NFT sensitivity to emission gamma-ray lines at 3$\sigma$ confidence level and an observation time $T = 10^5$~s. {\em Bottom panel}: Minimum Detectable Polarization of the \astena\ NFT and WFM-IS proposed telescopes compared with that of the AMEGO mission concept (see, e.g., \cite{Martinez22}), in the case of a 100 mCrab source for a measurement time of 100~ks at 3$\sigma$ confidence level. Adapted from \cite{Moita23}}.
\label{f:Astena-performance}
\end{figure}

\section{Prospects for Laue lens telescopes}
\label{s:Concluding}

From this review on the physics behind the Laue lenses background, the expectations from these focusing instruments and the results obtained thus far, some prescriptions for the future Laue lenses can be deduced:
\begin{itemize}
\item 
Bent crystals that give rise to curved diffracting planes and thus to a quasi-mosaic rocking curve, appear to be the best choice for maximizing the lens reflectivity;
\item 
The gluing of the crystals to substrates does not appear to be the best technology in order to achieve angular resolutions better than one arcmin. The best angular resolutions could be achieved by direct bonding with no use of glue. Other solutions, like the tilt mechanism of each crystal tile suggested, e.g.,  by  \cite{Lund21b}, appear complex for controlling a few thousand of crystals, in addition to its impact on overall lens transparency specially at lower energies, mass increase and filling factor of the lens.
\item 
Very high focal distances that require two satellites in formation flying have to be avoided for their increased complexity. 
\end{itemize}

SiLC-based Laue lenses appear a promising solution in the low energy band, given that the low atomic number of Silicon allows a  good diffraction efficiency  only up to 150--200 keV.

%
% Figure 22
%
%\begin{figure}[ht]
%\begin{center}
%\includegraphics[angle=0, width=0.40\textwidth]{Figures/fig-astena-inflight.jpg}
%\end{center}
%\vspace{-0.5cm}
%\caption{Artistic view of the ASTENA mission concept. The Laue lens (NFT) is at the top-center of the payload, while around NFT is the WFM-IS with a passband from 2~keV to 20~MeV.   Reprinted from \cite{Frontera21}.}
%\label{f:ASTENA}
%\end{figure}
%

As already mentioned (see, e.g., section~\ref{s:optical-properties}), in 2019 a Laue lens telescope, designed for satisfying the above prescriptions, was proposed as one of the two instruments (NFT $=$ Narrow Field Telescope), the other being a broad-band Wide Field Monitor Imager and Spectrometer (WFM-IS),  for the \astena\ mission concept 
%(see Fig.~\ref{f:ASTENA}) 
proposed in two white papers  to the European Space Agency (ESA) for its new long term programme "Voyage 2050" \cite{Frontera21,Guidorzi21}. The lens telescope is made of bent crystals of Si(111) and Ge(111) that cover the nominal energy band from 50 to 600 keV with a still significant effective area up to 700 keV. A description of the lens and its performance, in addition to some key science goals that can be faced with NFT, can be found in \cite{Frontera21}. Other science goals that could be faced with \astena\ can be found in \cite{Guidorzi21}.
The expected angular resolution and FOV of NFT have been given in section~\ref{s:optical-CDP}.
Assuming the position sensitive detector described in section~\ref{s:focalPSD}, the NFT expected sensitivity to continuum emission, to emission lines  and to detect polarization of celestial soft gamma--ray sources is shown in Fig.~\ref{f:Astena-performance}.  As far as the sensitivity to continuum emission, it is also shown  the comparison of NFT with other experiments and missions. Among them, it is also included the broad-band (0.15~MeV--3~GeV) gamma--ray  mission concept \eastrogam\ \cite{Deangelis18}, that, in the common energy range with \astena\ NFT, makes use of a Compton telescope \cite{Deangelis18}. As can be seen from the top panel of Fig.~\ref{f:Astena-performance}, in this range \astena\ NFT is about two orders of magnitude more sensitive, but with a narrow FOV (4 arcmin).  Thus NFT can perform deep studies of single sources and solve issues in which a high angular resolution is required, while \eastrogam\ is suitable for monitoring broad sky fields, achieving a much better sensitivity than the current gamma--ray instruments by integrating on year times the source observations.

Concerning NFT, the ongoing development of the elastic bending and direct bonding of super-polished crystals  to their substrates \cite{Mazzolari23} is expected to give a bright future to the Laue lens telescopes.

The next years will give an answer to the real capability of Laue lenses to successfully face the many open issues of the soft gamma--ray astronomy. There are several reasons to be optimistic.

\begin{acknowledgement}
First of all I wish to thank all past and current members of the High Energy Astrophysics Group of the Physics Department of the University of Ferrara and all other collaborators, not only in Ferrara, for their significant contribution to the Laue lens developments we have performed so far and their interest to the future projects in the field. I wish also to thank the members of the Universe editorial team for their useful suggestions, and our PhD student Lisa Ferro for her contribution to the figure preparation. Last, but not least, I wish to thank my wife for her patience in leaving me, very often, free from family needs. 

\end{acknowledgement}

%\bibliographystyle{spphys}
%bibliography{lens_biblio}
\end{document}